\documentclass[aps, prx, amsmath,amsfonts,reprint,superscriptaddress,longbibliography]{revtex4-1}
\usepackage{amsmath,amssymb,amsfonts}
\usepackage{color}
\usepackage{natbib}
\usepackage{graphicx}
\usepackage{dcolumn}
\usepackage{bm}
\usepackage{color,soul}

\usepackage[usenames]{xcolor}

\begin{document}

\title{Huygens' meta-devices for parametric waves}

\author{Mingkai Liu}
\email{mingkai.liu@anu.edu.au}
\affiliation {Nonlinear Physics Centre, Research School of Physics and Engineering, Australian National University, Canberra ACT 2601, Australia}
\author{David A. Powell}
\affiliation {Nonlinear Physics Centre, Research School of Physics and Engineering, Australian National University, Canberra ACT 2601, Australia}
\affiliation {School of Engineering and Information Technology, University of New South Wales, Canberra ACT 2610, Australia}
\author{Yair Zarate}
\author{Ilya V. Shadrivov}
\affiliation {Nonlinear Physics Centre, Research School of Physics and Engineering, Australian National University, Canberra ACT 2601, Australia}

\begin{abstract}
Huygens' metasurfaces have demonstrated almost arbitrary control over the shape of a scattered beam, however its spatial profile is typically fixed at fabrication time. Dynamic reconfiguration of this beam profile with tunable elements remains challenging, due to the need to maintain the Huygens' condition across the tuning range. In this work, we experimentally demonstrate that a time-varying meta-device which performs frequency conversion, can steer transmitted or reflected beams in an almost arbitrary manner, with fully dynamic control. Our time-varying Huygens' meta-device is made of both electric and magnetic meta-atoms with independently controlled modulation, and the phase of this modulation is imprinted on the scattered parametric waves, controlling their shapes and directions. We develop a theory which shows how the scattering directionality, phase and conversion efficiency of sidebands can be manipulated almost arbitrarily. We demonstrate novel effects including all-angle beam steering and frequency-multiplexed functionalities at microwave frequencies around 4 GHz, using varactor diodes as tunable elements. We believe that the concept can be extended to other frequency bands, enabling metasurfaces with arbitrary phase pattern that can be dynamically tuned over the complete 2$\pi$ range.

\end{abstract}


\maketitle

\section{Introduction}

Recent advances in Huygens' metasurfaces and meta-devices provide new insight into highly efficient wavefront shaping and scattering manipulation by incorporating electric and magnetic responses \cite{liu2010incident,pfeiffer2013metamaterial,monticone2013full,kim2014optical,decker2015high,asadchy2015broadband,chong2015polarization,epstein2016cavity,
kruk2016invited,PhysRevLett.117.256103,wong2016reflectionless,elsakka2016multifunctional,estakhri2016wave}, extending earlier studies of blazed gratings~\cite{magnusson_diffraction_1978}. While a large amount of work has been done on static Huygens' metasurfaces and meta-devices, advanced applications require that the performance of meta-devices can be tuned to adapt to varying operating conditions or requirements. Ideally, this would mean that the amplitude and phase response of a Huygens' meta-device can be tuned \emph{dynamically} in \emph{an arbitrary} fashion. Here, ``arbitrary'' means the amplitude and phase response can be \emph{independently tuned to all the possible states}; for example, to tune the phase response over a complete $2\pi$ range without changing the amplitude response, or to change the amplitude from zero to maximum while keeping the phase unchanged. While there have been some recent attempts to achieve independent tuning of Huygens' metasurfaces \cite{zhu2014dynamic}, fully arbitrary control remains very challenging to realize in practice, since truly arbitrary tuning requires independent control of not only the electric and magnetic responses, but also of the gain and loss.

To solve this problem, we propose parametric meta-devices, where some parameters of the structure can be dynamically modulated. When an electromagnetic wave with central frequency $\omega_0$ interacts with a system having properties modulated with frequency $\Omega\ll\omega_0$, new sideband frequencies $\omega_{n}=\omega_0+n\Omega\;$ ($n\in\mathbb{Z}$) are generated. The manifestation of this process can be found in systems of different scales, ranging from the energy level splitting of trapped cold atoms \cite{wineland1975proposed,RevModPhys.75.281}, Raman scattering of vibrational molecules \cite{shen1965theory,long1977raman,garcia1996collective}, Brillouin scattering due to opto-acoustic coupling \cite{brillouin1922diffusion,mandelstam1926light,chiao1964stimulated}, to the amplitude and frequency modulation of radio signals \cite{bakshi2009communication}, and the micro-Doppler effect widely employed in radar sensing \cite{chen2006micro,chen2011micro}. The manipulation and detection of these sidebands are of great importance from both fundamental and application points of view. A variety of novel effects explored recently in optical systems with dynamic modulation also rely on sideband control, such as sideband cooling \cite{wineland1975proposed,park2009resolved,chan2011laser,teufel2011sideband}, magnet-free optical isolation \cite{yu2009complete,PhysRevLett.109.033901,PhysRevLett.108.153901,estep2014magnetic}, and optomechanical interaction in the resolved sideband regime \cite{kippenberg2007cavity,aspelmeyer2014cavity}. 

Understanding and harnessing the effects of dynamic modulation in artificial subwavelegnth electromagnetic systems such as resonant particles and metasurfaces, could lead to various ultra-compact tunable devices and novel functionalities via the introduction of an additional degree of freedom -- time varying properties~\cite{liu2013self,liu2014spontaneous,hadad2015space,shaltout2015time, hadad2016breaking,shi2016dynamic,ShiYu2017OpticalCirculation,taravati2017nonreciprocal,taravati2017mixer}. One important feature of sideband generation in subwavelength systems is that the scattered phases of the sidebands have a gauge freedom and can be controlled by the dynamic modulation process. However, due to the symmetry of dipole scattering, a thin-layer of modulated elements scatters sidebands in both forward and backward directions. In addition, the limited modulation strength available in practice further suppresses the efficiency of energy conversion from the carrier wave to sidebands at the subwavelength scale.

 \begin{figure*}[t!]
\includegraphics[width=1.5\columnwidth]{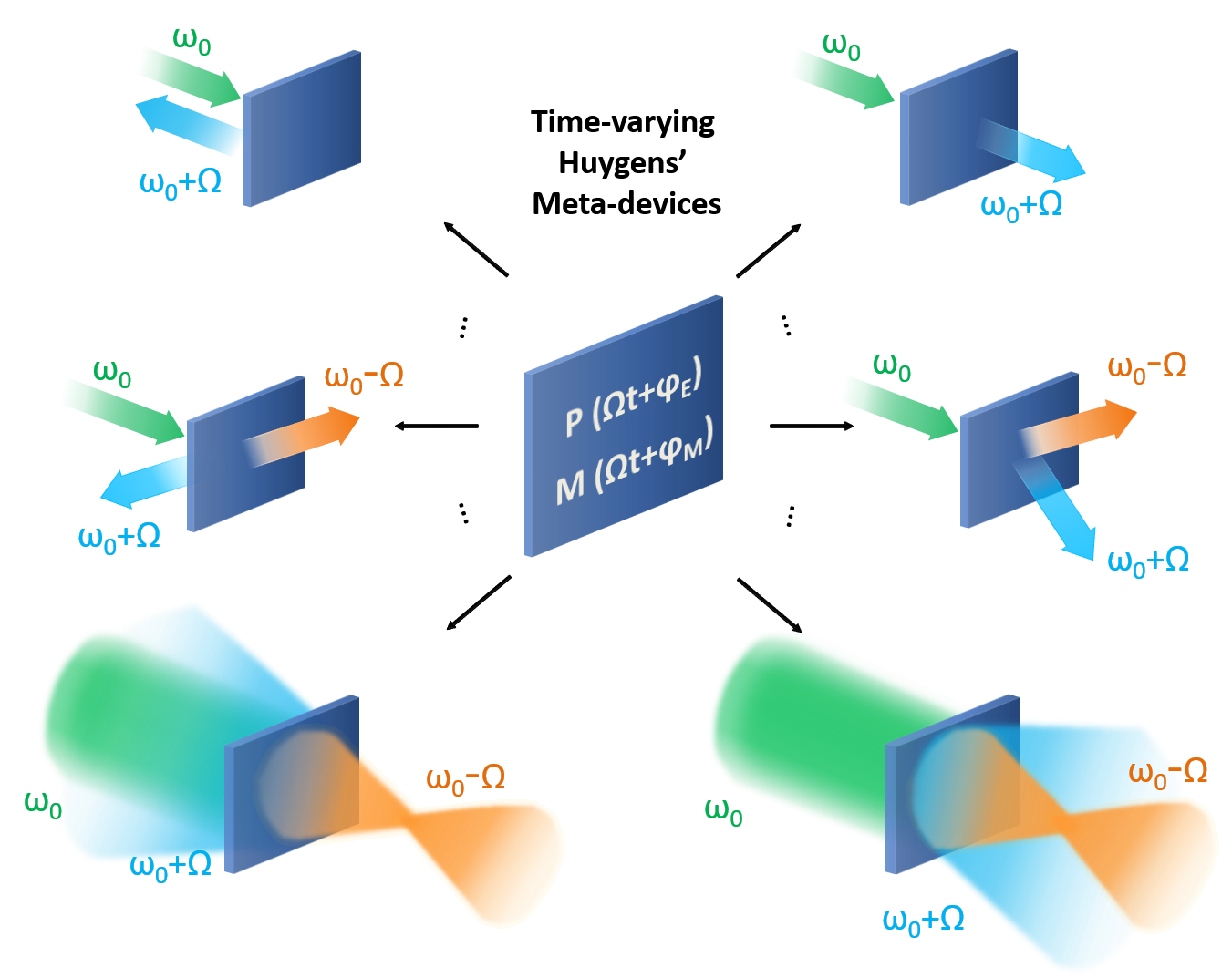}
\caption{ Schematic of time-varying Huygens' meta-devices for parametric waves. By independently modulating the electric and magnetic polarizations $P$ and $M$, the sideband scattering can be manipulated almost arbitrarily, and multiple functionalities can be achieved with the same Huygens' meta-device. (Top) Single sideband directive scattering. By changing the relative modulation phases $\varphi_{\rm E}$ and $\varphi_{\rm M}$, the directionality can be tuned between forward (top right) and backward (top left) mode. (Middle) Double sideband directive scattering, including double-sideband bidirectional scattering (middle left) and unidirectional scattering (middle right), where the two sidebands can be steered towards different directions via a linear phase gradient along the units of the meta-device. When a more complicated phase gradient is introduced, a time-varying Huygens' meta-device can function as different devices simultaneously for different sidebands. For example, under double-sideband bidirectional scattering mode, it can function as a concave mirror for one sideband and a convex lens for the other (bottom left); while under double-sideband forward scattering mode, it can function as lenses of opposite focusing properties for different sidebands (bottom right).  \label{fig:concept} }
\end{figure*}

Here, we develop and experimentally verify a theory of time-varying Huygens' meta-devices for parametric waves, which consist of electric and magnetic meta-atoms that can be modulated \emph{independently} via external stimuli. Compared to conventional static Huygens' meta-devices, these time-varying  Huygens' meta-devices do not aim to manipulate the carrier wave at frequency $\omega_0$, but instead efficiently convert the carrier wave into sideband frequencies $\omega_{n}$ via dynamic modulation (see Fig.~\ref{fig:concept}). Unlike static Huygens' meta-devices, where the amplitude and phase response to the carrier wave are determined by the intrinsic resonances of the meta-atoms, the sidebands scattered from a time-varying meta-device are fully controlled by the amplitude and phase of modulation.

Since the parities of electric and magnetic dipole radiation are preserved under modulation, the directivity of sideband scattering can be manipulated by the relative modulation phase of electric and magnetic meta-atoms, and high conversion efficiency from the carrier frequency to sidebands can be achieved within a layer of subwavelength thickness. By introducing a gradient of modulation phase along the unit cells of a metasurface, the sidebands can be steered in the transverse direction. We note that some recent studies already provided vivid numerical demonstrations of sideband generation and steering in time-varying Huygens' metasurfaces \cite{stewart2016finite,smy2017finite,vahabzadeh2017space}; in this study we reveal the physical mechanism and highlight the key components for achieving high conversion efficiency and full spatial control of parametric waves. We demonstrate this concept experimentally in the microwave frequency range, where we measure the sideband scattering of independently modulated electric and magnetic meta-atoms. Using optimized modulation signals, we achieve a high conversion efficiency of over $75\%$ from the carrier wave to the target  sidebands and successfully demonstrate various sideband scattering regimes with controlled directionality, including single-sideband unidirectional scattering  [Fig.~\ref{fig:concept} (top)], double-sideband unidirectional scattering  [Fig.~\ref{fig:concept} (middle right)], and double-sideband bi-directional scattering [Fig.~\ref{fig:concept} (middle left)]. Using a finite time-varying Huygens' meta-device, we demonstrate novel effects including all-angle (360-degree) beam steering and frequency-multiplexed functionalities [Fig.~\ref{fig:concept} (bottom)]. These results set the foundation of more advanced Huygens' meta-devices for parametric waves.

\section{Analytical model} \label{sec:theory}

\subsection{General description based on the effective impedance model} \label{sec:theory_general}

The physics of time-varying Huygens' meta-devices can be described by the boundary conditions of a thin sheet with space-time-dependent electric and magnetic polarizations $\mathbf{P}(\mathbf{r},t)$ and $\mathbf{M}(\mathbf{r},t) $, dubbed the generalized sheet transition conditions (GSTCs)~\cite{idemen1987boundary}:
 \begin{eqnarray}\label{eq:pfullbound1}
\mathbf{n}\times\left(\mathbf{H}_{\Vert}^{\rm f}+\mathbf{H}_{\Vert}^{\rm b}-\mathbf{H}_{\Vert}^{\rm i}\right)&=&\frac{{\rm d}}{{\rm d}t}\mathbf{P_{\Vert}},\\
\mathbf{n}\times\left(\mathbf{E}_{\Vert}^{\rm f}-\mathbf{E}_{\Vert}^{\rm b}-\mathbf{E}_{\Vert}^{\rm i}\right)&=&-\frac{{\rm d}}{{\rm d}t}\mathbf{M_{\Vert}},
\label{eq:pfullbound2}
\end{eqnarray}
where $\mathbf{n}$ is the normal vector of the surface; the superscripts `$\rm{i}$', `$\rm{f}$' and `$\rm{b}$' represent the incident, forward scattered and backward scattered fields, respectively; `${\Vert}$' denotes the components parallel to the surface. The more general form of GSTCs also includes the normal components of polarizations $\mathbf{P_{\bot}}$ and $\mathbf{M_{\bot}}$ (see Appendix \ref{sec:boundcon}), but in this work, we limit our discussion to a meta-device with polarizations only in the lateral directions  ($\mathbf{P_{\bot}}=\mathbf{M_{\bot}}=0$). 

When a slow periodic modulation with a fundamental frequency of $\Omega$ is introduced in the polarizations, i.e. $\mathbf{P_{\Vert}}=\mathbf{P_{\Vert}}(\mathbf{r},\Omega t)$ and $\mathbf{M_{\Vert}}=\mathbf{M_{\Vert}}(\mathbf{r},\Omega t)$, all the time-dependent components of Eqs.~(\ref{eq:pfullbound1}) and (\ref{eq:pfullbound2}) can be decomposed in the frequency domain in the form of $F(\mathbf{r},t)=\sum_{n=-\infty}^{+\infty}F_{n}(\mathbf{r})e^{-{\rm i}\omega_{n}t}$, where $\omega_{n}=\omega_{0}+n\Omega$ are the sideband frequencies \cite{bakshi2009communication},  with $\omega_0$ the carrier frequency of the incident field. The electric and magnetic polarizations can be achieved using time-varying meta-atoms with effective Fourier components of electric and magnetic dipole moments $\mathbf{p}_{n}=\mathbf{P}_{n}A$, and $\mathbf{m}_{n}=\mathbf{M}_{n}A/\mu$, respectively, where A is the area of the unit-cell, and $\mu$ is the permeability of the medium surrounding the meta-device. The boundary equations (\ref{eq:pfullbound1}) and (\ref{eq:pfullbound2}) for sideband $\omega_n$ are then given by
\begin{eqnarray}\label{eq:boundn1}
\mathbf{n}\times\left(\mathbf{H}_{n,\Vert}^{\rm f}+\mathbf{H}_{n,\Vert}^{\rm b}-\delta_{n0}\mathbf{H}_{n,\Vert}^{\rm i}\right)&=&-\mathrm{i}\omega_n\mathbf{p}_{n,\Vert}/A,\\
\mathbf{n}\times\left(\mathbf{E}_{n,\Vert}^{\rm f}-\mathbf{E}_{n,\Vert}^{\rm b}-\delta_{n0}\mathbf{E}_{n,\Vert}^{\rm i}\right)&=&\mathrm{i}\omega_n\mu\mathbf{m}_{n,\Vert}/A.\label{eq:boundn2}
\end{eqnarray}
$\delta_{n0}$ is the Kronecker delta, which is $1$ for $n=0$, and $0$ otherwise. It is clear that the space-time dependency of the scattered fields on the two sides of a meta-device is determined by the space-time variation of the effective dipole moments; only in the case of static meta-devices ($\Omega=0$) is the frequency of the scattered fields identical to the incident one. 

However, the Fourier components of the effective dipole moments are not independent but interact with each other via parametric modulation. In order to describe this parametric process in the meta-device, we extend our previous models for static meta-devices~\cite{liu2012optical,PhysRevB.90.075108}. Generally, the interaction between the electric (magnetic) meta-atom at position $\bf r$ and the incident wave can be expressed in a compact form: $Z_{\rm E(M)}I_{\rm E(M)}=V_{\rm E(M)}$, where $Z_{\rm E(M)}$ is the effective impedance of the meta-atom in the array with all the mutual interaction taken into account (see Appendix \ref{sec:impmatrix} for details); $I_{\rm E(M)}$ is the mode amplitude (current amplitude); $V_{\rm E(M)}$ is the effective electromotive force evaluated by the overlap of the incident field and the mode profile~\cite{liu2012optical,PhysRevB.90.075108}. When a periodic modulation $f(\Omega t-\varphi)$ is introduced to the meta-atoms, $Z$, $I$ and $V$ become time-dependent, and they can be expanded in the frequency domain by a Fourier series. Representing each meta-atom as a series RLC circuit, the input electromotive force is equal to the total voltage at any point in time, which requires that the coefficients of the Fourier decomposition of $Z(t)$, $I(t)$ and $V(t)$ satisfy the following equation (see Appendix \ref{sec:impmatrix} for details)
\begin{eqnarray}\label{eq:impedanceFourier}
 \sum_{m=-\infty}^{+\infty}Z_l^{(m)}I_{m}e^{-{\rm i}(l+m)(\Omega t-\varphi)}=V_{n}e^{-{\rm i}n(\Omega t-\varphi)}.
\end{eqnarray}
Here, $l=n-m$. The subscripts denote the order of the Fourier coefficients of the time-varying $Z(t)$, $I(t)$ and $V(t)$, while the superscript in $Z_l^{(m)}$ denotes that the Fourier coefficient is calculated at frequency $\omega_m$. $\varphi$ is the relative phase introduced from the modulation, which is effectively a time-delayed replica of the modulation waveform. The dynamics of the modulated meta-atom at position $\bf r$ can then be expressed in a compact matrix form:
\begin{eqnarray}\label{eq:intmatrix}
\overleftrightarrow{\mathbf{Z}_{\rm eff}}\mathbf{I}	=	\mathbf{V}, 
\end{eqnarray}
where
\begin{eqnarray}\label{eq:impedance}
\overleftrightarrow{\mathbf{Z}_{\rm eff}}&=&\left[\begin{array}{ccccc}
\ddots & \vdots &  & \vdots\\
\cdots & Z_{0}^{(m)} & \cdots & Z_{m-n}^{(n)} e^{{\rm i}\varphi(m-n)} &\cdots\\
 & \vdots & \ddots & \vdots\\
\cdots & Z_{n-m}^{(m)}e^{{\rm i}\varphi(n-m)} & \cdots & Z_{0}^{(n)} & \cdots\\
 & \vdots &  & \vdots & \ddots
\end{array}\right].\nonumber\\\\
\mathbf{I}&=&\left[\begin{array}{c}
\cdots, I_{m}e^{{\rm i}m\varphi},\cdots,I_{n}e^{{\rm i}n\varphi},\cdots
\end{array}\right]^T,\\
\mathbf{V}&=&\left[\begin{array}{c}
\cdots, V_{m}e^{{\rm i}m\varphi}, \cdots, V_{n}e^{{\rm i}n\varphi}, \cdots
\end{array}\right]^T. 
\end{eqnarray} 
The zeroth order impedance $Z_0^{(m)}$ describes the linear response of the meta-atom at frequency $\omega_m$, where the effect of mutual interaction with other meta-atoms has been taken into account; the $(m-n)_{\rm th}$ order component $Z_{m-n}^{(n)}$ characterizes the frequency conversion from $\omega_m$ to $\omega_n$, which is originated from the dynamic modulation of the self-impedance.  For more details regarding the construction of Eq.~(\ref{eq:impedance}), see Appendix \ref{sec:impmatrix}.

Under normally incident plane wave excitation $\mathbf{E}^{\rm i}$, the Fourier component of the effective electromotive force acting on the electric and magnetic meta-atoms  can be expressed as
\begin{eqnarray}
V_{\mathrm E,n}&=&\int\mathbf{j}_{{\rm E},n}\cdot\mathbf{E}^{\rm i}{\rm d}^{3}\mathbf{r}=u_{\mathrm E,n} E^{\rm i}, \label{eq:VE}\\
V_{\mathrm M,n}&=&\int\mathbf{j}_{{\rm M},n}\cdot\mathbf{E}^{\rm i}{\rm d}^{3}\mathbf{r}={\rm i}k_0 E^{\rm i} u_{\mathrm M,n},\label{eq:VM}
\end{eqnarray}
$u_{{\rm E},n}=-\mathrm{i}\omega_n p_n/I_{{\rm E},n}$ and $u_{{\rm M},n}=m_n/I_{{\rm M},n}$ are the Fourier components of the normalized effective electric and magnetic dipole moments, which can be defined by the Fourier component of the normalized current distribution ${\bf j}_n$ \cite{liu2012optical,PhysRevB.90.075108} (see Appendix \ref{sec:impmatrix} for more details). Solving Eq.~(\ref{eq:intmatrix}) independently for both electric and magnetic meta-atoms, we can find the current amplitudes $I_{{\rm E (M)},n}$, then the dipole moments at frequency $\omega_n$ can be readily found as
\begin{eqnarray}\label{eq:edipole}
p_{n}(\mathbf{r})&=&{\rm i}I_{{\rm E},n}(\mathbf{r})u_{{\rm E}}e^{{\rm i} n\varphi_{\rm E}(\mathbf{r})}/\omega_{n},\\
m_{n}(\mathbf{r})&=&I_{{\rm M},n}(\mathbf{r})u_{{\rm M}}e^{{\rm i} n\varphi_{\rm M}(\mathbf{r})}. \label{eq:mdipole}
\end{eqnarray}
Here, $\varphi_{\rm E(M)}$ is the relative phase introduced from the modulation of the electric (magnetic) meta-atom.

Once we substitute Eqs.~(\ref{eq:edipole}) and (\ref{eq:mdipole}) into Eqs.~(\ref{eq:boundn1}) and (\ref{eq:boundn2}), it becomes clear that an additional phase of $n\varphi$ is introduced to the sideband fields ($E_n, H_n$) via modulation. This feature of time-varying Huygens' meta-devices offers a unique opportunity in controlling wave scattering -- since the modulation phase $\varphi$ can be chosen arbitrarily for different unit cells, one can generate any desired spatial phase pattern along the meta-device simply by modifying the local modulation phase $\varphi(\mathbf{r})$. As a result, the phase of each unit cell can be dynamically tuned over a complete $2\pi$ range, giving our structure advantage over existing static Huygens' meta-devices that rely on the deliberate overlap and balance of electric and magnetic resonances. Moreover, unlike conventional sideband scattering processes, the functionality of time-varying Huygens' meta-devices can be actively tuned by changing the modulation phase difference between the electric and magnetic meta-atoms. 

\subsection{Modulation with uniform amplitude and a linear phase gradient}\label{sec:theory_example1}

To give an illustrative example, we consider the simple case of a homogeneous metasurface consisting of identical units, where the response of the meta-atoms is polarization-independent. The incident carrier wave propagates in the normal direction, and the dynamic modulation of the meta-atoms (of the same type) has identical amplitude but a periodic linear phase gradient: $\varphi(y)=\varphi_0+Gy$, with $G$ being the spatial frequency of modulation and $\varphi_0$ being the modulation phase at $y=0$. As a result, the solution of the system is a series of Floquet modes characterized by ($\omega_n, \beta_n$)
\begin{eqnarray}
\mathbf{I}(y)&=&\left[\begin{array}{c}
\cdots, I_{n}(y)e^{{\rm i}(n\varphi_0+\beta_n y)},\cdots
\end{array}\right]^T,
\end{eqnarray}
where each sideband frequency $\omega_n$ is scattered with a transverse wave vector  $\beta_n=\beta_0+nG$.

When the modulation frequency $\Omega$ is much smaller than the linewidth of the resonance, the difference between the impedance evaluated at $\omega_l$ and $\omega_m$ is negligible, and the effective impedance matrix in Eq.~(\ref{eq:impedance}) can be simplified as (for details, see Appendix \ref{sec:impmatrix}):
\begin{eqnarray} \label{eq:imp_ntoeplitz}
\overleftrightarrow{\mathbf{Z}_{\rm eff}}&=&\left[\begin{array}{ccccc}
Z_{\beta,-n} & \cdots & Z_{-n}e^{-{\rm i}n\varphi} & \cdots & Z_{-2n}e^{-{\rm i}2n\varphi}\\
\vdots & \ddots & \vdots & \ddots & \vdots\\
Z_{n}e^{{\rm i}n\varphi} & \cdots & Z_{\beta,0} & \cdots & Z_{-n}e^{-{\rm i}n\varphi}\\
\vdots & \ddots & \vdots & \ddots & \vdots\\
Z_{2n}e^{{\rm i}2n\varphi} & \cdots & Z_{n}e^{{\rm i}n\varphi} & \cdots & Z_{\beta,n}
\end{array}\right]\nonumber\\
\end{eqnarray}
$Z_{\beta,n}$ is the zeroth order effective impedance of Floquet mode ($\omega_n, \beta_n$).  For a normally incident carrier wave ($\beta_0=0$), the sidebands $\omega_{n}$ and $\omega_{-n}$ have opposite transverse wave vectors: $\beta_n=-\beta_{-n}$, i.e. they are deflected towards opposite transverse directions, and their impedances can be approximated as identical: $Z_{\beta,n}=Z_{\beta,-n}$. In the special situation of uniform modulation without any phase gradient ($G=0$), all the sidebands are scattered in the normal direction ($\beta_n=0$), and the effective impedance can be considered the same for all sidebands: $Z_{\beta,n}=Z_{\beta,m}$, in this special situation, Eq.~(\ref{eq:imp_ntoeplitz}) becomes a Toeplitz matrix [Eq.~(\ref{eq:imp_toeplitz}) in Appendix \ref{sec:impmatrix}].
  
We further assume that the change of the mode profile is negligible during modulation (i.e. ${\bf j}_n=0$ for $n\neq 0$), and only the zeroth order terms remain in the effective electromotive force $\bf V$, as well as in the normalized dipole moments $u_{\rm E}$ and $u_{\rm M}$ of Eq.~(\ref{eq:VE}) and (\ref{eq:VM}), then
\begin{eqnarray}
\mathbf{V}&=&\left[\begin{array}{c}
V_{-n},\cdots,V_{0},\cdots,V_{n}\end{array}\right]^{T},
\end{eqnarray}
with $V_{n}=V\delta_{n0}$. These approximations work well in the adiabatic limit of modulation, and the sideband spectra calculated based on the impedance model converge and agree well with the full-wave simulation when the order of truncation increases. See Appendix \ref{sec:imp} for details.

In order to give a simple explicit expression, we truncate both the sideband and the modulation to the first order; for purely reactive modulation, we define $Z_{1}=Z_{-1}=-{\rm i}\xi$, where $\xi\in\mathbb{R} $ (see Appendix \ref{sec:imp} for more discussion). The effective impedance and the electromotive force at position $y$ are simplified as 

\begin{eqnarray} \label{eq:impFloq2}
\overleftrightarrow{\mathbf{Z}_{\rm eff}}&=&\left[\begin{array}{ccc}
Z_{\beta} & -\text{{\rm i}}\xi e^{-{\rm i}\varphi} & 0\\
-\text{{\rm i}}\xi e^{{\rm i}\varphi} & Z_{0} & -\text{{\rm i}}\xi e^{-{\rm i}\varphi}\\
0 & -\text{{\rm i}}\xi e^{{\rm i}\varphi} & Z_{\beta}
\end{array}\right], \\
\mathbf{V}&=&\left[\begin{array}{c}
0, V, 0
\end{array}\right]^T.
 \end{eqnarray}
$Z_{0}=Z_{\beta,0} $ and $Z_{\beta}=Z_{\beta,\pm 1}$ are the effective impedances for the carrier wave and the first order sidebands.
The mode amplitudes can be expressed in a compact form:
\begin{eqnarray}\label{eq:current2}
\mathbf{I}=\left[\begin{array}{c}
I_{-1}e^{{\rm -i\varphi}}\\
I_{0}\\
I_{1}e^{{\rm i\varphi}}
\end{array}\right]=\frac{V}{\sigma}\left[\begin{array}{c}
{\rm i}\xi e^{{\rm -i\varphi}}\\
Z_{0}\\
{\rm i}\xi e^{{\rm i\varphi}}
\end{array}\right]
 \end{eqnarray}
where $\sigma=Z_{0}Z_{\beta}+2\xi^{2}$. Equation~(\ref{eq:current2}) shows that while the zeroth order response $I_0$ is only affected by the modulation amplitude $\xi$, both the amplitude $\xi$ and the modulation phase $\varphi$ play a key role at the sideband frequencies. 

For a particular Floquet mode, the effective impedance of the meta-atom can be characterized with $Z={-\rm i}X+R^{(\rm rad)}+R^{(\rm ohm)}$, where $X$, $R^{(\rm ohm)}$ and $R^{(\rm rad)}$ correspond to frequency-dependent reactance, ohmic loss and radiative loss, respectively.  From the passivity condition, the following relations for electric and magnetic meta-atoms hold for the TE polarized Floquet mode ($\omega_n, \beta_n$) (see Appendix \ref{sec:radloss} for details): 
\begin{eqnarray}\label{eq:radloss}
R_{{\rm E},n}^{(\rm rad)}=\frac{\eta u_{\rm E}^{2}k_n}{2A\kappa_n},\,\, R_{{\rm M},n}^{(\rm rad)}=\frac{\eta u_{\rm M}^{2}k_n\kappa_n}{2A}.
\end{eqnarray}
$\eta$ is the wave impedance of the surrounding space; $k_n=\omega_n/c$ is the total wave vector and $\kappa_n=\sqrt{k_n^2-\beta_n^2}$ is the longitudinal component. Applying Eq.~(\ref{eq:current2}) for both electric and magnetic meta-atoms via Eqs.~(\ref{eq:edipole}) and (\ref{eq:mdipole}), and using the relations in Eqs.~(\ref{eq:VE}), (\ref{eq:VM}) and (\ref{eq:radloss}), we can derive the dipole moments $p_n$ and $m_n$, as well as the forward and backward fields $E_{n}^{{\rm f}}$ and $ E_{n}^{{\rm b}}$ via Eqs.~(\ref{eq:boundn1}) and (\ref{eq:boundn2}) [for details see Eqs.~(\ref{eq:scatfx}) and (\ref{eq:scatbx})]. The generalized scattering parameters for TE polarization can be defined as ${\rm r}_n=\sqrt{\kappa_n/k_n}E_{n}^{{\rm b}}/E^{{\rm i}}$ and ${\rm t}_n=\sqrt{\kappa_n/k_n}E_{n}^{{\rm f}}/E^{{\rm i}}$, where the factor $\sqrt{\kappa_{n}/k_{n}}$ originates from the local power conservation \cite{epstein2014passive} (see Appendix \ref{sec:radloss} for more details). The final expressions are given by

\begin{eqnarray} \label{eq:r0_f}
\rm{r}_{0}&=&-\frac{R_{{\rm E},0}^{({\rm rad})}Z_{{\rm E,\beta}}}{\sigma_{{\rm E}}}+\frac{R_{{\rm M},0}^{({\rm rad})}Z_{{\rm M,\beta}}}{\sigma_{{\rm M}}},\\
\rm{t}_{0}&=&1-\frac{R_{{\rm E},0}^{({\rm rad})}Z_{{\rm E},\beta}}{\sigma_{{\rm E}}}-\frac{R_{{\rm M},0}^{({\rm rad})}Z_{{\rm M,\beta}}}{\sigma_{{\rm M}}}.\label{eq:t0_f}\\
\rm{r}_{\pm1}&=&-\sqrt{\frac{\kappa_{\pm 1}}{k_{\pm1}}}\frac{{\rm i}R_{{\rm E},\beta}^{({\rm rad})}\xi_{{\rm E}}e^{{\rm \pm i\varphi_{{\rm E}}}}}{\sigma_{{\rm E}}}+\sqrt{\frac{k_{\pm1}}{\kappa_{\pm1}}}\frac{{\rm i}R_{{\rm M},\beta}^{({\rm rad})}\xi_{{\rm M}}e^{{\rm \pm i\varphi_{{\rm M}}}}}{\sigma_{{\rm M}}},\label{eq:r1_f}\nonumber\\\\
\rm{t}_{\pm1}&=&-\sqrt{\frac{\kappa_{\pm1}}{k_{\pm1}}}\frac{{\rm i}R_{{\rm E},\beta}^{({\rm rad})}\xi_{{\rm E}}e^{{\rm \pm i\varphi_{{\rm E}}}}}{\sigma_{{\rm E}}}-\sqrt{\frac{k_{\pm1}}{\kappa_{\pm1}}}\frac{{\rm i}R_{{\rm M},\beta}^{({\rm rad})}\xi_{{\rm M}}e^{{\rm \pm i\varphi_{{\rm M}}}}}{\sigma_{{\rm M}}}.\label{eq:t1_f}\nonumber\\
\end{eqnarray}

To have frequency conversion with maximal efficiency (limited only by the fraction of energy dissipated in the meta-atoms), no energy may be transmitted or reflected at the carrier frequency, i.e. $\rm{t}_{0}=\rm{r}_{0}=0$. Applying this condition to Eqs.~(\ref{eq:r0_f}) and (\ref{eq:t0_f}), we get the requirement for the modulation term:  $\xi^{2}=\frac{1}{2}Z_{\beta}(2R^{({\rm rad})}_0-Z_{0})$, which should be satisfied independently for electric and magnetic meta-atoms. It is clear from Eqs.~(\ref{eq:r0_f}) and (\ref{eq:t0_f}) that due to the symmetry of the dipole scattered field, perfect conversion is impossible if only electric or magnetic dipole response is employed. For a lossless metasurface, $R^{(\rm ohm)}=0$, the condition for complete frequency conversion can be simplified as $|\xi|^2=\frac{1}{\sqrt{2}}Z_{0}^*Z_{\beta}$, which can be strictly satisfied when $Z_{0}$ and $Z_{\beta}$ have the same phase angle. From the relations in Eq.~(\ref{eq:radloss}), this implies that $Z_{{\rm E,}\beta}=\frac{k_{n}}{\kappa_{n}}Z_{{\rm E,}0},
  Z_{{\rm M,}\beta}=\frac{\kappa_{n}}{k_{n}}Z_{{\rm M,}0}$; as a result, the required impedance modulation $|\xi_{{\rm E}}|^{2}=\frac{k_{n}}{2\kappa_{n}}|Z_{E,0}|^{2},|\xi_{{\rm M}}|^{2}=\frac{\kappa_{n}}{2k_{n}}|Z_{M,0}|^{2}$. The corresponding sideband scattering parameters in Eqs.~(\ref{eq:r1_f}) and (\ref{eq:t1_f}) are then simplified as
\begin{eqnarray} 
\rm{r}_{\pm 1}&=& \frac{-\rm i}{2\sqrt{2}}e^{{\rm i}(\psi_{\rm E}\pm\varphi_{\rm E})}[1-e^{{\rm i}(\Delta\psi\pm\Delta\varphi)}], \label{eq:r1_lossless}\\
\rm{t}_{\pm 1} &=& \frac{-\rm i}{2\sqrt{2}}e^{{\rm i}(\psi_{\rm E}\pm\varphi_{\rm E})}[1+e^{{\rm i}(\Delta\psi\pm\Delta\varphi)}].\label{eq:t1_lossless}
\end{eqnarray}
$\psi_{\rm E(M)}=\mathrm{Arg}(|Z_{\rm E(M),0}|/Z_{\rm E(M),0})$ is the intrinsic linear phase response of the electric (magnetic) meta-atoms and $\Delta\psi=\psi_{\rm M}-\psi_{\rm E}$ is the intrinsic phase difference; $\Delta\varphi=\varphi_{\rm M}-\varphi_{\rm E}$ is the phase difference between the modulation signals applied to the electric and magnetic meta-atoms. 

Equations~(\ref{eq:r1_lossless}) and (\ref{eq:t1_lossless}) highlight the capability of time-varying Huygens' meta-devices to control the directionality of sidebands by changing the modulation phase difference between electric and magnetic meta-atoms. For example, for overlapping electric and magnetic resonances ($\Delta\psi=0$), $|\rm{r}_{\pm 1}|=0$ and $|\rm{t}_{\pm 1}|=1/\sqrt{2}$ when the relative modulation phase $\Delta\varphi=0$; while $|\rm{r}_{\pm 1}|=1/\sqrt{2}$ and $|\rm{t}_{\pm 1}|=0$ when $\Delta\varphi=180^{\circ}$. This indicates that a single time-varying Huygens' meta-surface can function as either a transmissive (see the middle-right panel of Fig.~\ref{fig:concept}) or reflective device simply by changing the relative phase $\Delta\varphi$. Another example is when $\Delta\psi= \Delta\varphi=\pm 90^{\circ}$, the $+1_{\mathrm st}$ and $-1_{\mathrm st}$ orders can be well separated in opposite directions: $\rm |r_{-1}|=|t_{+1}|=0, |t_{-1}|= |r_{+1}|=1/\sqrt{2}$ (see the middle-left panel of Fig.~\ref{fig:concept}). These effects will be demonstrated experimentally in Sec.~\ref{sec:exp} and highlighted in Figs.~\ref{fig:experiment_dyn} and~\ref{fig:experiment_dyn2}.

In contrast to static metasurfaces, overlapped and balanced electric and magnetic resonances at the carrier frequency is not always required in time-varying Huygens' meta-devices, for two reasons: 1) different types of directive sideband scattering require different detuning between the electric and magnetic resonances; 2) the unbalanced linear response of electric and magnetic meta-atoms can be compensated in the parametric process by adjusting the amplitudes, phases and waveforms of the dynamic modulation of electric and magnetic meta-atoms. 

We emphasize that while the above discussion provides some useful insight, the condition for perfect conversion shown above is based on the approximation where both the impedance modulation and the sidebands can be truncated to the first order. In practice, however, this approximation is difficult to achieve since sideband generation is a cascaded process. Moreover, impedance modulation of resonators generally changes both the amplitude and the phase of the carrier wave, and the phase changes nonlinearly with the modulation signal, which inevitably introduces high order sidebands. This effect becomes more pronounced when strong modulation is applied. In order to maximize the energy conversion to the first order sidebands, we need to introduce higher order correction terms in the modulation waveform to suppress the undesirable high order sidebands, as will be shown below. 

\section{Design of Huygens' units for parametric waves}\label{sec:designHU}

To validate the concept, we design electric and magnetic meta-atoms working in the microwave regime using full-wave simulation (CST Microwave studio). As a first step, we study a Huygens' unit (a pair of electric and magnetic meta-atoms) in a metallic rectangular waveguide, which is easier to characterize experimentally. We emphasize that while the theoretical discussion above focused on a meta-device excited by a normally-incident plane wave, the impedance model is very general and can be easily extended to other situations. Here, the pair of closely spaced resonators are positioned at the center of the waveguide and are excited by the fundamental waveguiding mode. Although the effective impedance $Z$ of the meta-atom in a rectangular waveguide is different from the one in a periodic array, the whole system is still closely related to the situation discussed in Sec.~\ref{sec:theory_example1}, since in both cases  the scattered fields from the electric and magnetic meta-atoms have opposite parities, and the waveguide system can be considered as a special situation discussed in Sec.~\ref{sec:theory_example1}, in which the carrier wave and the sidebands share the same scattering channels (forward and backward in the normal direction). Therefore, we expect that the features discussed in Sec.~\ref{sec:theory_example1} can also be observed in this system. Indeed, we find that even a basic Huygens' unit made of a pair of electric and magnetic meta-atoms is sufficient to demonstrate the effect of directional sideband scattering in conjunction with a high conversion efficiency. 

\begin{figure}[t!]
\includegraphics[width=1\columnwidth]{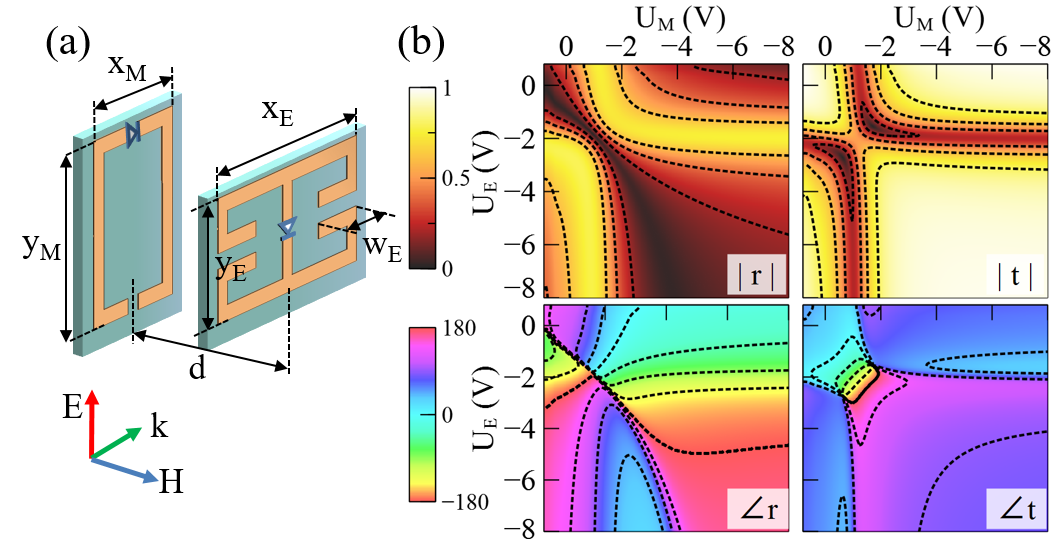}
\caption{(a) Design of a Huygens' unit element (sizes in mm): $x_{\rm M}=7.87,\,y_{\rm M}=12, x_{\rm E}=14.25,\,y_{\rm E}=8,\, w_{\rm E}=3.89,\, d=6.5$. The track width and gap size are both 1 mm. Measured at 3.7 GHz ($\lambda\approx$ 81 mm), the overall size of the Huygens' unit is subwavelength: $x_{\rm E}\approx 0.18\lambda,  y_{\rm M}\approx 0.15\lambda, d\approx 0.08\lambda$. The thickness of the substrates is 0.4 mm. (b) The simulated transmission and reflection coefficients at $\omega_0=$3.7 GHz as a function of the DC bias voltages ($U_{\rm E}, U_{\rm M}$), in the absence of dynamic modulation. \label{fig:numerical_design} }
\end{figure} 

 \begin{figure}[t!]
\includegraphics[width=1\columnwidth]{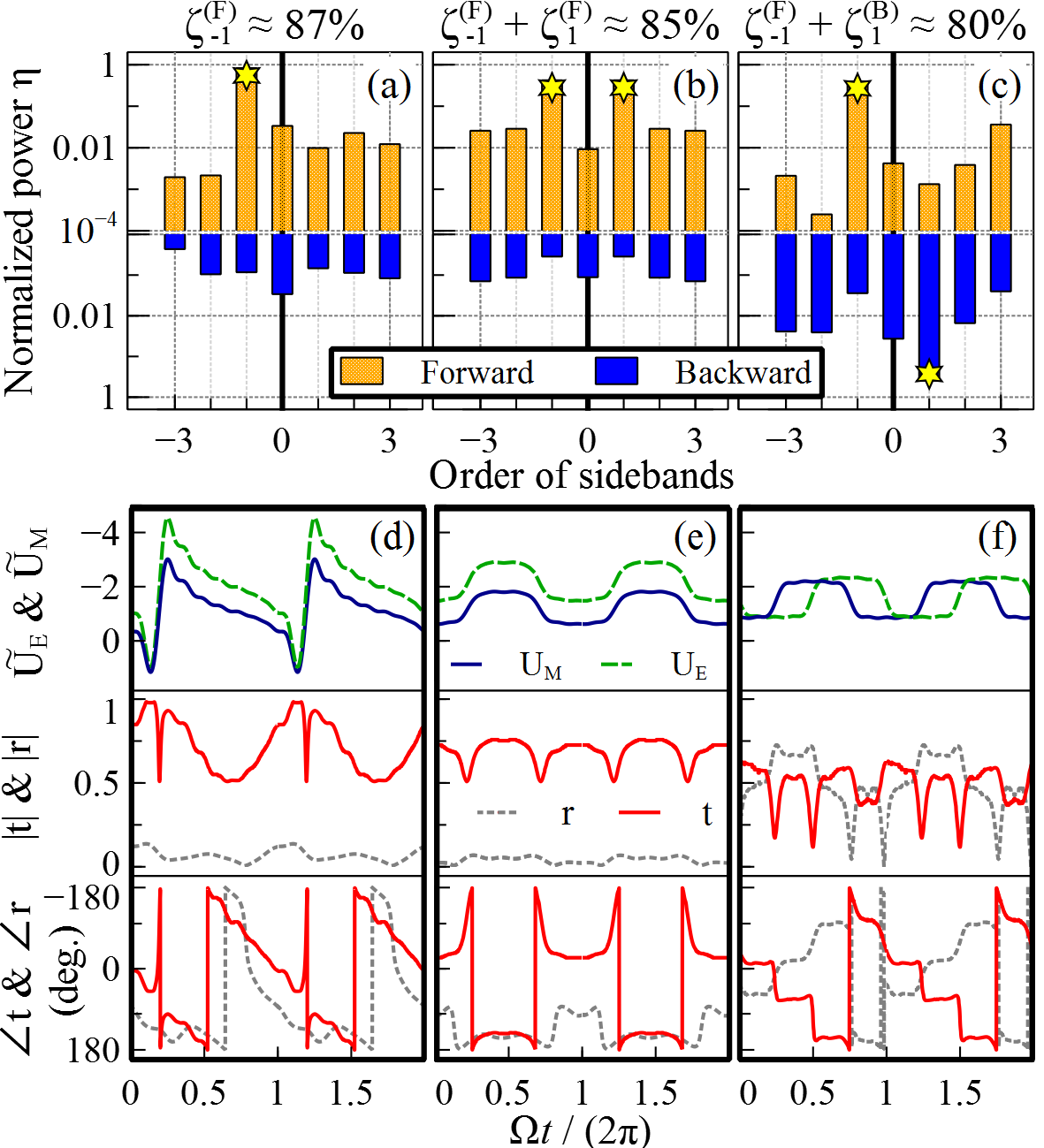}
\caption{ Simulated normalized sideband spectra for forward and backward scattering under the optimized modulation waveforms for (a) single-sideband directive forward scattering, (b) double-sideband directive forward scattering, and (c) double sideband bidirectional scattering. The spectra are normalized to the total scattered power of all sidebands up to the tenth order, including the carrier frequency; $\zeta_{n}^{(\rm F/B)}$ is the normalized scattered power for order $n$ in the forward/backward direction. For clarity, only sidebands up-to the third orders, with a normalized power larger than  $10^{-4}$ are shown. Stars indicate the dominant sidebands. The corresponding modulation voltage signals, as well as the resulting modulation of transmission and reflection, are plotted in (d), (e) and (f).  \label{fig:numerical} }
\end{figure} 

The Huygens' unit consists of an electric and a magnetic split-ring resonators, which are printed individually on a Rogers RO4003 substrate
($\epsilon_r$ = 3.5, loss tangent 0.0027, substrate thickness 0.4 mm), and the whole unit is positioned at the center of a WR229 rectangular waveguide that supports a TE$_{10}$ mode [see geometry and parameters of meta-atoms in the caption of Fig.~\ref{fig:numerical_design} (a)]. We use varactor diodes (SMV1405) as the voltage-tunable capacitors. Under appropriate DC bias voltages, the electric and magnetic resonances overlap. Figure \ref{fig:numerical_design} (b) shows the simulated linear response of the unit at $\omega_0/(2\pi)=$3.7 GHz as a function of the DC bias voltages ($U_{\rm E}$, $U_{\rm M}$). There exists a parameter regime around $U_{\rm E}/U_{\rm M}\approx 2$, where the reflection approaches zero and the transmission phase changes by around $2\pi$, which is a typical manifestation of the Huygens' condition. 

We are interested in the behavior of the unit around the overlapped resonance when dynamic modulation is introduced. We choose the carrier frequency to be $\omega_0/(2\pi)=$3.7 GHz, and assume that the modulation frequency $\Omega$ is sufficiently low such that the sidebands produced by the dynamic modulation can be calculated in the adiabatic limit (see Appendix \ref{sec:num_sim} for more discussion), where we approximate the time-varying transmission and reflection signals with voltage-dependent stationary linear responses (see Appendix \ref{sec:num_sim} for more discussion). In this paper, we only study the behavior in the adiabatic limit of modulation, while a detailed theoretical discussion of modulation beyond the adiabatic limit can be found in Ref.~\cite{Momchil2017Exact}. 

The modulation voltage is defined as $\tilde{U}_{\rm E (M)}(t)=U_{\rm E (M), amp}f(t)+U_{\rm E (M), offs}$, with $U_{\rm E (M), amp}$, and $U_{\rm E (M), offs}$ being the modulation amplitude and the DC offset; $f(t)=\sum_{l=1}^{N}a_{\rm E (M)}^{(l)}\cos[l(\Omega t-\varphi_{\rm E (M)})]+b_{\rm E (M)}^{(l)}\sin[l(\Omega t-\varphi_{\rm E (M)})]$ is the normalized waveform constructed using a Fourier series. As has been noted in Sec.~\ref{sec:theory_example1}, we introduced high order terms in the modulation waveform to suppress the undesirable higher order sidebands, and we truncate the highest order to $N=8$ since higher order terms ($N>8$) only bring in negligible improvement. For each set of modulation signals [$\tilde{U}_{\rm E}(t)$, $\tilde{U}_{\rm M}(t)$], we can obtain the modulated scattering parameters at each time step via a one-to-one mapping of the voltage-dependent  stationary linear response from Fig.~\ref{fig:numerical_design} (b), and the corresponding sidebands can be calculated via Fourier transformation (see Appendix \ref{sec:num_sim} for more details).

We employ a genetic algorithm to optimize the waveform in order to maximize both the directivity and the power of the chosen sidebands (see Appendix Sec.~\ref{sec:GA} for details). As an example, Figs.~\ref{fig:numerical} (a) to (c) depict the normalized spectra for three different types of directive sideband scattering. The spectra are normalized to the \emph{total scattered power of all sidebands including the carrier frequency}, and more than 80\% of the scattered power is converted to the targeted sidebands. The corresponding modulation waveforms and the resulting modulation in transmission and reflection are plotted in Figure~\ref{fig:numerical} (d) to (f). To simplify the optimization procedure, we use the same normalized waveform for $\tilde{U}_{\rm E}$ and $\tilde{U}_{\rm M}$. This restriction is not essential, however our empirical tests indicated no improvement in efficiency when allowing these waveforms to differ. The optimized coefficients of the modulating signals are detailed in Table~\ref{table:theory} of Appendix \ref{sec:GA}. 

The results of the optimization process reveal that different types of sideband generation require very different modulation waveforms. To achieve a high conversion efficiency for single-sideband forward scattering [Fig.~\ref{fig:numerical} (a)], the transmission coefficient is ideally modulated linearly over $360^{\circ}$ (see Appendix \ref{sec:ideal} for more discussion); the optimized result indeed shows that the phase of transmission changes almost linearly over $360^{\circ}$, and the required modulation waveform has a highly asymmetric sawtooth-like shape [Fig.~\ref{fig:numerical} (d)]. In contrast, for double-sideband unidirectional forward scattering [Fig.~\ref{fig:numerical} (b)], the transmission is ideally amplitude modulated by a purely sinusoidal waveform (see Appendix \ref{sec:ideal}), and thus the optimized $U_{\rm E}$ and $U_{\rm M}$ have a sinusoidal-like waveform and are modulated in phase ($\Delta\varphi\approx 0$) [Fig.~\ref{fig:numerical} (e)]. Note that the reflection remains low during the modulation cycle and thus the  transmission amplitude modulation is achieved via the absorption modulation of the meta-device. The most demanding situation is double-sideband bidirectional scattering [Fig.~\ref{fig:numerical} (c)], which ideally requires that both the transmission and reflection phases are modulated linearly over $360^{\circ}$, but with opposite signs of the slopes (see Appendix \ref{sec:ideal}). This challenging requirement can be maximally satisfied using electric and magnetic meta-atoms simultaneously since we can find a parameter regime from Fig.~\ref{fig:numerical_design} (b) where the transmission and reflection phases change dramatically. The final optimized result from Fig.~\ref{fig:numerical} (f) confirms that the transmission phase is indeed modulated over $360^{\circ}$ with a positive slope while the reflection phase is modulated with a negative slope; $U_{\rm E}$ and $U_{\rm M}$ are modulated with a phase-lag $\Delta\varphi\approx -90^{\circ}$, indicating there is an intrinsic phase difference of $\Delta\psi\approx -90^{\circ}$ between the electric and magnetic response. 

These results are consistent with our previous discussion based on Eqs.~(\ref{eq:r1_lossless}) and (\ref{eq:t1_lossless}), which shows the $\Delta\varphi$ required for different types of sideband control. In fact, the optimized sideband spectra  can also be reproduced using the impedance model introduced in Sec.~\ref{sec:theory_example1} by taking into account the higher order terms in the impedance matrix shown in Eq.~(\ref{eq:imp_toeplitz}). To evaluate the accuracy of the impedance model, we calculate the Fourier coefficients $Z_{{\rm E(M)},n}$ using the time-varying scattering coefficients shown in Figs.~\ref{fig:numerical} (d) to (f), and apply these Fourier coefficients in the impedance model to calculate the mode amplitudes $I_{{\rm E(M)},n}$ as well as the sideband spectra (see Appendix \ref{sec:imp} for details). The results clearly show that as the order of truncation increases, the spectra calculated with the impedance model converge and agree well with the full-wave simulation. In fact, the maximum relative error of the dominant sidebands is already below 5\% when we truncate the harmonics to the first order sidebands (see Fig.~\ref{fig:imp_compare} in Appendix \ref{sec:imp}). Note that different from the simplified case discussed in Eq.~(\ref{eq:impFloq2}), the second order terms $Z_{\pm 2}$ are also preserved in Eq.~(\ref{eq:imp_toeplitz}), which allows us to calculate the case where the modulation waveform is asymmetric, such as single-sideband conversion.

\section{Experimental realization}\label{sec:exp}

\subsection{A Huygens' unit in a rectangular waveguide}

 \begin{figure}[t!]
\includegraphics[width=1\columnwidth]{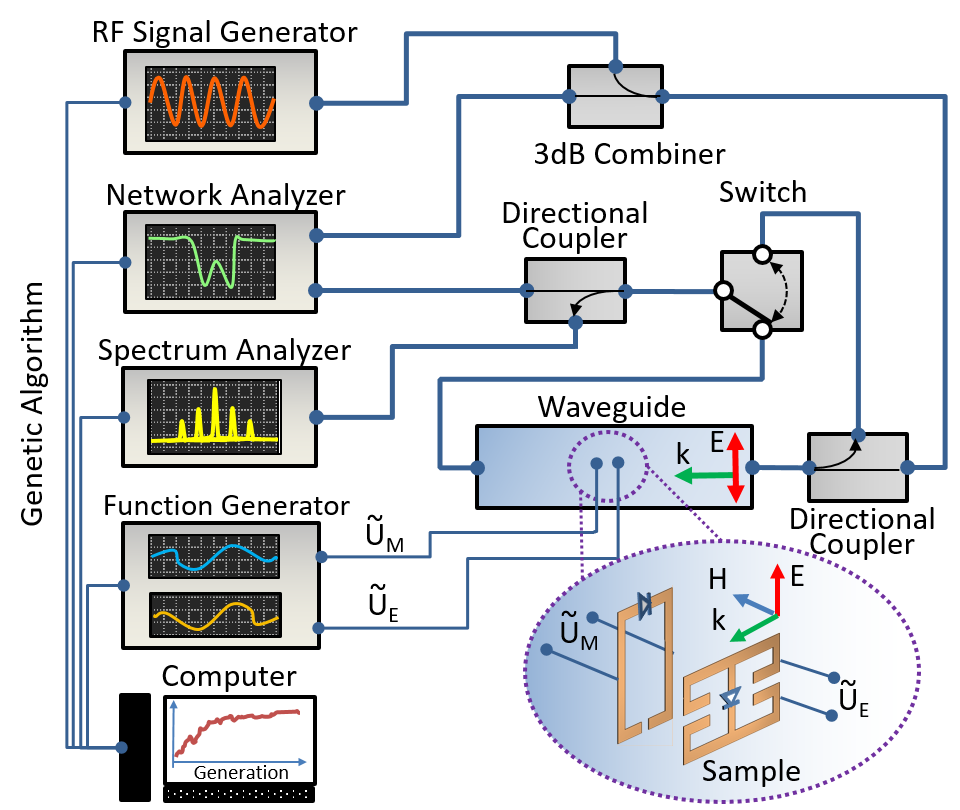}
\caption{Schematic of the experimental microwave setup. RF Signal Generator: HP-8673B; Vector Network Analyzer: Rohde \& Schwarz ZVB-20; Spectrum Analyzer: Rohde \& Schwarz FSV-30; Arbitrary Function Generator: Tektronix AFG 3022B; Switch: Agilent 8763B coaxial switch. For clarity, the substrates of the meta-atoms are not shown.
\label{fig:setup} }
\end{figure} 

\begin{figure}[t!]
\includegraphics[width=1\columnwidth]{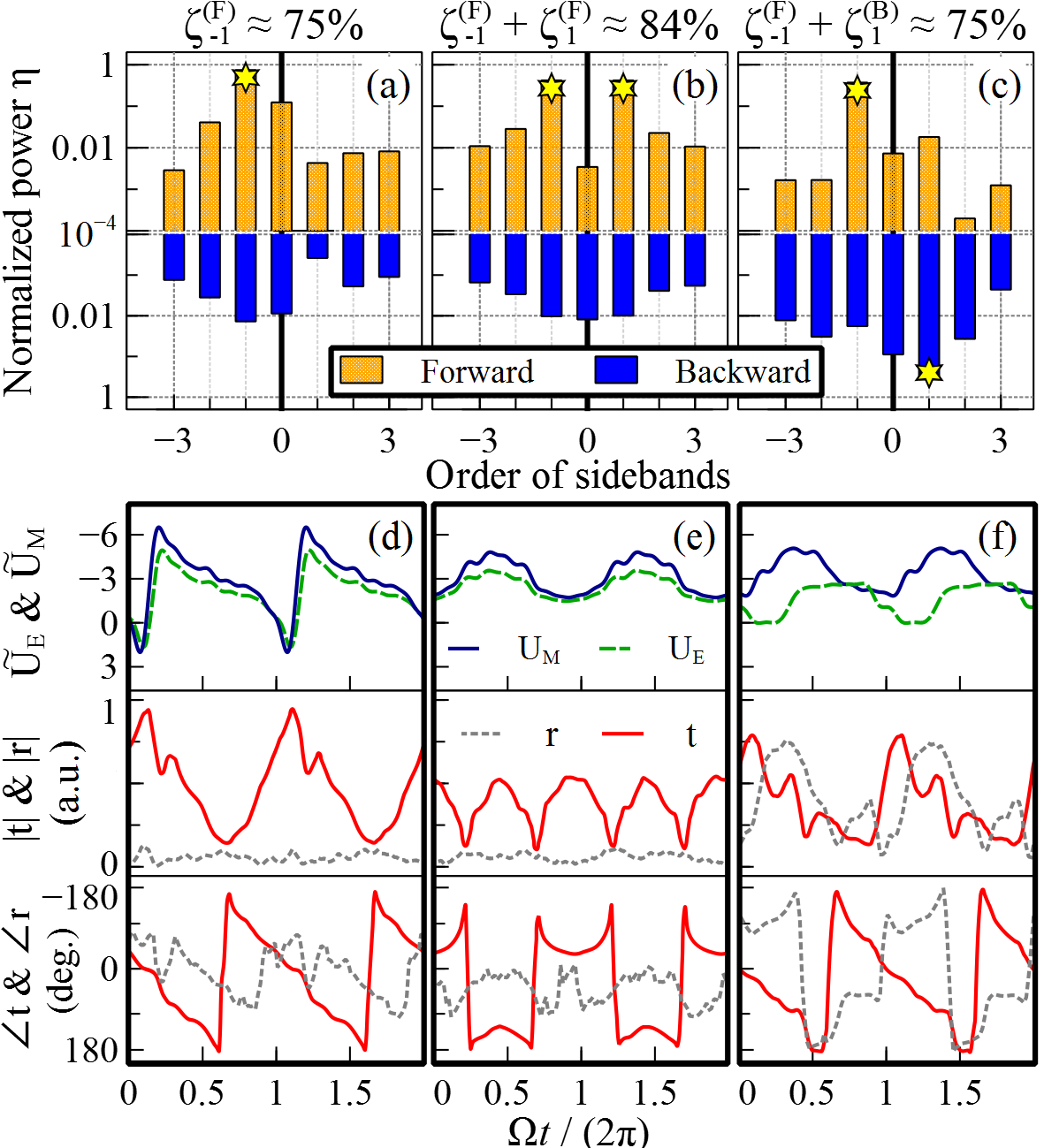}
\caption{Measured normalized sideband spectra of a pair of electric and magnetic meta-atoms in a rectangular waveguide. The spectra are normalized to the total scattered power of all sidebands up to the tenth order. (a) single-sideband forward scattering, (b) double-sideband forward scattering, and (c) double-sideband bi-directional scattering. Stars indicate the dominant sidebands. (d) to (f) are the corresponding modulation waveforms and the time-varying transmission and reflection signals. \label{fig:experiment} }
\end{figure} 

\begin{figure}[t!]
\includegraphics[width=1\columnwidth]{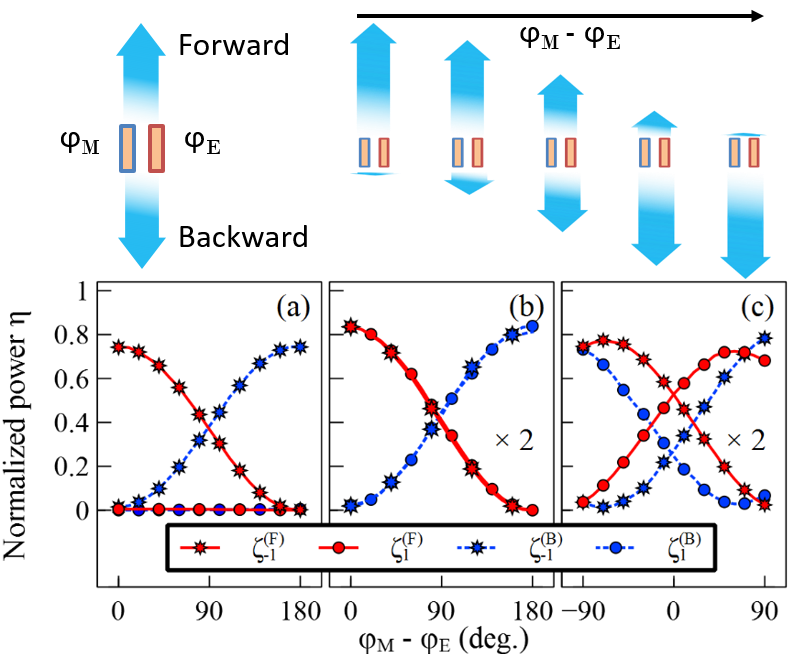}
\caption{(top) Schematic representation of the dynamic control of sideband scattering in the rectangular waveguide, where the directionality of sidebands can be tuned by the modulation phase difference $\varphi_{\rm M}-\varphi_{\rm E}$. (a) to (c) Measured normalized scattered power of the first order sidebands as a function of the modulation phase difference for three different types of directional sideband scattering. (a) Single-sideband unidirectional scattering; (b) double-sideband unidirectional scattering; (c) double-sideband bidirectional scattering. The sideband powers are normalized to the total scattered power of all sidebands (including the carrier frequency). For clarity, the curves are scaled by a factor of 2 in figures (b) and (c). }\label{fig:experiment_dyn} 
\end{figure}

To demonstrate the dynamic control of sideband scattering, we fabricated the electric and magnetic meta-atoms separately on a Rogers RO4003 printed circuit board; varactor diodes were soldered into the additional gaps in the meta-atoms as voltage-tunable elements. The schematic of the microwave setup is shown in Fig.~\ref{fig:setup}. The pair of meta-atoms were fixed on a polystyrene foam holder (with a relative permittivity close to 1) forming a Huygens' unit, and positioned at the center of a rectangular waveguide (see Fig.~\ref{fig:setup}). We first employed the network analyzer to measure the resonance spectrum of the sample and set the DC bias of the arbitrary function generator to tune the two meta-atoms such that their resonances overlap. The resonant frequencies of the fabricated meta-atoms red-shift compared to the simulation, due to the slight difference in substrate permittivity and the connection to the bias network; nevertheless, the resonant frequencies are still in the designed range and can be easily tuned to overlap. When the electric and magnetic resonances are separated, two reflection peaks were observed; as the DC bias voltages $(U_{\rm E,offs}$, $U_{\rm M,offs})\approx(-2.0 V, -2.5 V)$, the resonances of the pair overlap, with a reflection dip observed at around 3.64 GHz and the measured transmission phase varies over $2\pi$, which is a clear evidence for the Huygens condition. We set this point as the initial point for optimization and introduced a microwave CW carrier signal with $\omega_0/(2\pi)=3.64$ GHz from the RF signal generator. Periodic modulation signals with a base frequency $\Omega/(2\pi)=2$ MHz were generated by the arbitrary function generator to modulate the meta-atoms; the interaction between the carrier wave and the modulated meta-atoms generate sidebands on both sides of the carrier frequency, which were measured by the spectrum analyzer.

The modulation voltage signals were constructed using the same definition of Fourier series shown in Sec.~\ref{sec:designHU}. The offset voltage $U_{\rm E(M),offs}$, the amplitude $U_{\rm E(M),amp}$, the phase $\varphi_{\rm E(M)}$ as well as the coefficients of the waveforms were optimized using a genetic algorithm to maximize the power conversion to the desirable sidebands (see Appendix \ref{sec:GA} for details). Figure~\ref{fig:experiment} (a), (b) and (c) show the normalized scattering spectra for three different types of directive sideband scattering; the normalized power of the dominant sidebands is around 75\% for single-sideband unidirectional scattering and double-sideband bidirectional scattering, and around 84\% for double-sideband unidirectional scattering. The corresponding modulation waveforms and the time-varying transmission and reflection signals (measured with I/Q mode of the spectrum analyzer) are shown in  Fig.~\ref{fig:experiment} (d), (e) and (f). All the important features of time-varying amplitude and phase match well the theoretical predictions shown in the Fig.~\ref{fig:numerical} (d) to (f). The optimized coefficients of the modulation signals are listed from Table~{\ref{table:SSF}} to Table~{\ref{table:DSB}} in Appendix \ref{sec:GA}. These results confirm that, with appropriate modulation waveforms, a Huygens' design can achieve various directional sideband scattering with high conversion efficiency.   

Further, to demonstrate the capability to tune the directionality dynamically, we kept the optimized waveforms unchanged and only tuned the relative modulation phase of electric and magnetic meta-atoms. Remarkably, for all three types of directive sideband scattering, the directionality of the sideband scattering can be tuned dynamically in between the forward and backward mode, without sacrificing the total conversion efficiency [Fig.~\ref{fig:experiment_dyn} (a), (b) and (c)]. This property is particularly attractive since a time-varying Huygens' meta-device can function as a lens (transmissive mode), a mirror (reflective mode), or something in between, without requiring additional optimization of the waveform but a simple change of the relative phase of modulation. The sinusoidal-like change of the power with respect to $\Delta\varphi=\varphi_{\rm M}-\varphi_{\rm E}$ is also consistent with the prediction from Eqs.~(\ref{eq:r1_lossless}) and (\ref{eq:t1_lossless}): $|r_{\pm 1}|^2\propto 1-\cos(\Delta\psi\pm\Delta\varphi)$ and $|t_{\pm 1}|^2\propto 1+\cos(\Delta\psi\pm\Delta\varphi)$. 

\subsection{A  finite Huygens' meta-device in a parallel-plate waveguide}

\begin{figure*}[t!]
\includegraphics[width=1.7\columnwidth]{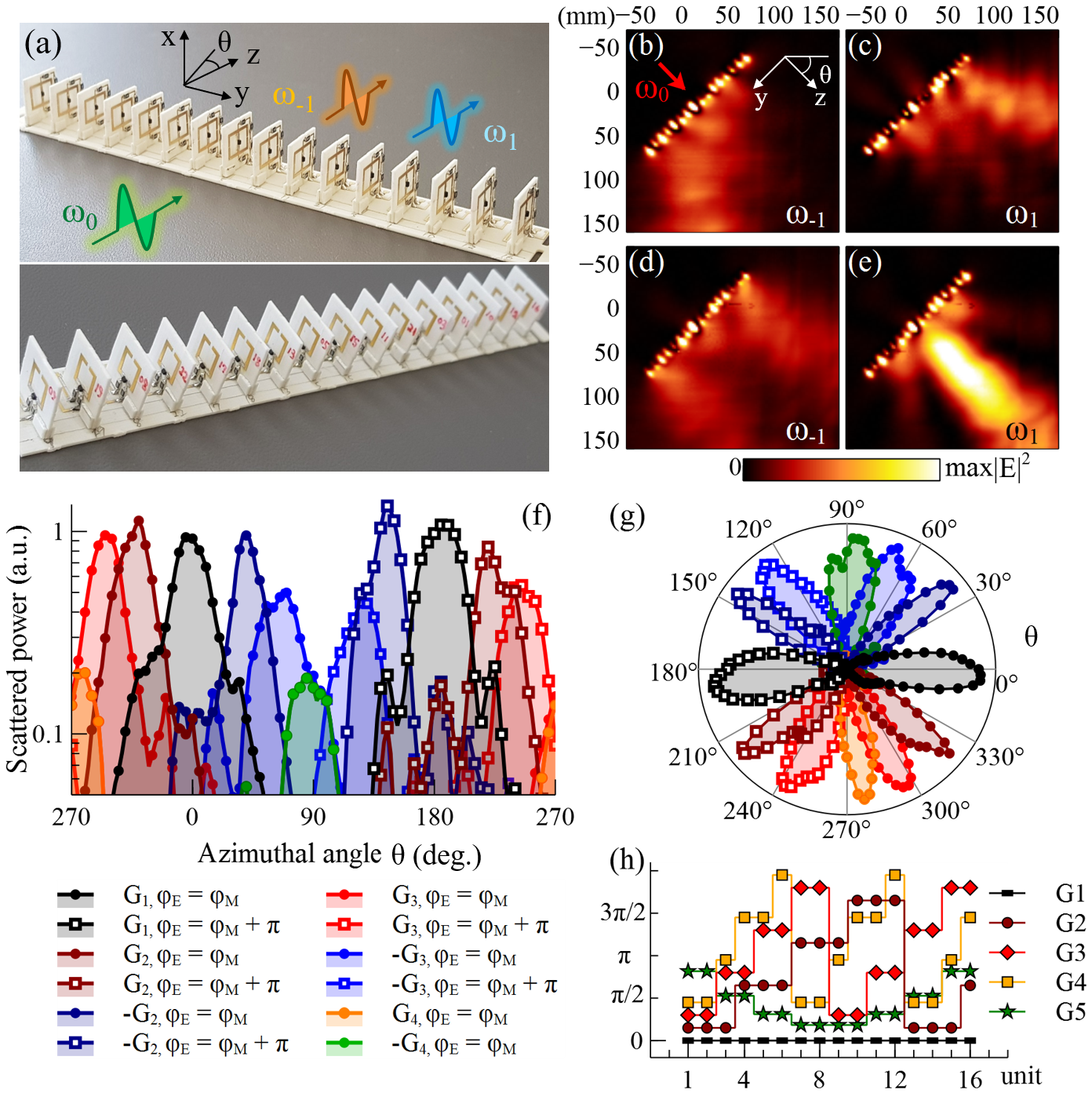}
\caption{(a) Photographs of the fabricated time-varying Huygens' metadevice. The electric meta-atoms (top) and magnetic meta-atoms (bottom) are fabricated on the opposite sides of the substrate. For detailed geometries, see Fig.~\ref{fig:array_unit} of Appendix \ref{sec:num_sim}. (b) and (c) Measured scattered field intensity at sideband frequencies $\omega_{\pm 1}$ under the modulation phase pattern G3 [see (f)], showing simultaneous beam-steering of the two side-bands in different directions. (d) and (e)  Measured scattered field intensity under phase pattern G5, showing simultaneous focusing and defocusing of the two sidebands. (f) Measured angular distributions of the scattered power at sideband $\omega_{-1}$  under different modulation phase patterns. The plots are normalized to the peak power of plot ``$\mathrm{G_1}, \varphi_{\mathrm E}=\varphi_{\mathrm M}$'', i.e., under a uniform modulation phase and in the forward scattering mode. The plots are shown in a logarithmic scale in the y-direction, and normalized power level below 0.05 (-13 dB) is not shown for clarity. (g) Polar plots of the angular distribution of sideband scattered power. For clarity,  each plot has been normalized to its peak power and has been shown in a linear scale along the radial direction; sidelobes below 0.05 (-13 dB) are not shown.  (h) Schematic of the modulation phase patterns. \label{fig:experiment_dyn2} }
\end{figure*} 

One advantage of time-varying Huygens' meta-devices is that they allow dynamic complex wavefront control in ways impossible using conventional linear tunable metasurfaces, such as all-angle beam steering and frequency-multiplexed functionalities. To demonstrate these effects, we designed and fabricated an array of Huygens' units working inside a two-dimensional field scanner based on a parallel-plate waveguide, as shown in Fig.~\ref{fig:experiment_dyn2} (a) (for the detailed designed geometries, see Fig.~\ref{fig:array_unit} of Appendix \ref{sec:num_sim}). The electric and magnetic meta-atoms are printed on the opposite sides of a single substrate to enhance the mechanical stability.

 The array is composed of 16 identical units, each consisting of an electric and a magnetic meta-atom that can be modulated independently. To mimic normally incident plane wave excitation, we generated a collimated beam using a parabolic mirror to direct the signals from a monopole source antenna.  We employed four dual channel arbitrary function generators to introduce 8 channels of independent modulation, where four of them were applied to the electric meta-atoms, and the other four used to modulate the magnetic meta-atoms. For the details of the experimental setup, see Fig.~\ref{fig:2dwaveguide_setup} and \ref{fig:2dwaveguide} and the corresponding discussion in Appendix \ref{sec:Expmeas}.

 To demonstrate dynamic metasurface functionality, we first tuned the bias voltages such that the two resonances overlap around 4 GHz. Then we introduced a CW carrier signal at $\omega_0/(2\pi)=$4 GHz and modulated the array with 4 synchronized arbitrary function generators. The modulation waveforms were first optimized to achieve double-sideband forward scattering for sidebands $\omega_{\pm 1}$ under a uniform modulation phase pattern G1 [see Fig.~\ref{fig:experiment_dyn2} (h)]; the optimized coefficients of the modulation signals are listed in Table~{\ref{table:DSB_array}} in Appendix \ref{sec:GA}. For simplicity, we only introduced different modulation phases at different units while keeping the modulation waveforms of all the electric or magnetic meta-atoms identical (see Fig.~\ref{fig:experiment_dyn2} (h) for the phase patterns). Due to the limited number of independent modulation channels, we only generated phase patterns with 4 discrete phase levels.

 Since different sidebands experience different imprinted phases, the time-varying meta-device allows frequency-multiplexed functionalities. When the first order term is dominant in the modulation waveforms, the phase imprinted on sidebands $\omega_{\pm 1}$ can be considered as conjugated. We confirmed this unique feature by mapping the two-dimensional distribution of the field intensity under the modulation phase patterns G3 and G5, as shown in Fig.~\ref{fig:experiment_dyn2} (b) to (e). The phase pattern G3 allows large angle beam steering ($>$60 degrees) of sidebands $\omega_{-1}$ and $\omega_{1}$ towards different directions, which was further confirmed by the scattering pattern in Fig.~\ref{fig:experiment_dyn2} (f). A more interesting effect occurred when the phase pattern G5 was applied -- the time-varying meta-device functioned as a convex lens for $\omega_{1}$ and a concave lens for $\omega_{-1}$ at the same time. By properly engineering the modulation waveforms, additional functionalities can be multiplexed into higher order sidebands.

 We measured the angular distribution of sideband power via a circular scan around the center of the sample. By introducing different modulation phase patterns, we achieved directional beam steering of the sidebands over the whole 360$^\circ$. Figure~\ref{fig:experiment_dyn2} (f) depicts the measured angle-dependent power for sideband $\omega_{-1}$. For clarity, we normalize each measurement and plot them in Fig.~\ref{fig:experiment_dyn2} (g). As has been shown in Fig.~\ref{fig:experiment_dyn}, adding an additional $\pi$ difference between the modulation phases $\varphi_{\mathrm E}$ and $\varphi_{\mathrm M}$ allows the beam to be steered from the forward (marked by circles) to the backward (marked by squares) direction. Such an all-angle directional beam steering highlights the unique capability of time-varying meta-devices as it is a highly nontrivial task for conventional tunable metasurfaces or beam deflectors.

 The asymmetry of the measured scattered power in Fig.~\ref{fig:experiment_dyn2} (f) under conjugated phases originates from the inhomogeneous response of the Huygens' units. This was mainly introduced during the hand-soldering of the varactor diodes and resistors, and it can be overcame by employing a more sophisticated fabrication process. We also note that the peak power reduces at large beam-steering angle, which can be attributed to two main reasons. The first one is due to the fact that we kept the modulation waveforms unchanged during beam steering. In fact, due to the mutual interaction among the units, the effective impedance of the Floquet mode will have a noticeable change at large steering angles compared to the scattering in the normal direction. We expect that additional improvement of the conversion efficiency can be obtained if the modulation waveforms are optimized for each steering angle. The second reason is the reduced directivity due to the decreased effective aperture at large steering angle. In fact, this is a common problem in all array antenna systems \cite{Balanis2016antenna}, and the directivity can be further improved by increasing the sample size (currently $\sim 2\lambda$) and adapting more sophisticated designs of the meta-atoms.

\section{Discussion and conclusion}\label{sec:discussion}
The concept of time-varying Huygens' meta-devices studied here could provide new tools and possibilities for research and applications that generate and manipulate sidebands. The ability to achieve a high conversion efficiency of sidebands is particularly attractive since it could enable various ultra-compact devices. One potential example is compact isolators. A naive design could consist of a bandpass filter that only transmits $\omega_{0}$ at the input side, followed by a time-varying Huygens' meta-device that achieves single sideband up-conversion ($\omega_0 \rightarrow \omega_1$), and a bandpass filter that only transmits $\omega_{1}$ at the output side. Since the process of sideband generation is nonreciprocal, a wave propagating in the reverse direction with frequency $\omega_{1}$ would be further up-converted to $\omega_2$ after passing the time-varying Huygens' meta-device, and be blocked by the filter at the input side. 

Another potential application based on the effects shown in Fig.~\ref{fig:experiment_dyn2} is a compact beam deflector that could steer one or more sidebands at different directions and scan them over the full $4\pi$ solid angle. This functionality could benefit the development of new compact radar systems for full-angle and multi-target detection. For example, existing frequency-modulated continuous wave (FMCW) radars widely employed in vehicles for obstacle detection can only “see” things within a certain angle range, and thus multiple radars are required to cover 360 degrees of view. This solution might become too costly and bulky when it is applied to light-weight platforms such as drones and small robots. A naive vision is to utilise a time-varying metasurface as the transmitter of the radar, from which different sidebands can be steered to different directions. The angular directions of the obstacles all around can be detected simultaneously with a single monopole receiver by identifying the frequencies of the signals bounced back.

The proof-of-concept design presented in this paper shows a promising start, yet further optimization and even new designs are required to facilitate implementation in more practical areas and in a more scalable fashion. We note that while the relatively high quality factor of the meta-atoms employed in the current design allows large modulation with a small voltage ($<$ 10 V), the peak absorption of the system can reach 65\% when the electric and magnetic resonances overlap. From simulation we notice that the percentages of the energy lost in the varactor diodes, metal tracks and substrates are around 66\%, 20\% and 14\%, respectively. To reduce the absorption loss, one can design meta-atoms with a lower Q factor by increasing the scattering loss and avoiding the self-resonance of the diode. In addition, one can employ tuning mechanisms that have lower loss, such as MEMS/NEMS \cite{chicherin2011analog, zheludev2012metamaterials, liu2013dynamic,han2014mems}, tunable capacitors based on ferroelectric thin films \cite{tagantsev2003ferroelectric,sazegar2012beam}, or transistors.

While the demonstrated dynamic sideband control is limited to the simple case of a one-dimensional array with only 8 independent channels,  this is not a fundamental limitation. Using an FPGA or micro-controller to achieve independent modulation over a large two-dimensional array is technically possible, as has been shown recently in the static tuning of a Huygens' meta-device \cite{chen2017reconfigurable}. The extension of the idea to terahertz and even optical frequencies is feasible by using other high-speed modulation mechanisms, such as electro-optical \cite{ju2011graphene, lira2012electrically, lee2012switching}, optomechanical/acousto-optical \cite{li2014photonic, liu2016polarization, zheludev2016reconfigurable} and nonlinear optical effects  \cite{lapine2014colloquium,minovich2015functional,shcherbakov2015ultrafast,shcherbakov2017ultrafast}.

To conclude,  we introduced the concept of time-varying Huygens' meta-devices and studied the sideband generation and manipulation in this type of system. We showed both theoretically and numerically that dynamic modulation of Huygens' meta-devices provides unique opportunities in manipulating parametric wave scattering. Importantly, we designed and fabricated prototype meta-devices working at microwave frequencies, and successfully demonstrated controlled generation and directive scattering in the experiment, with a high conversion efficiency ($>75\%$) from the carrier wave to the target sidebands. We also demonstrated novel effects that are difficult to achieve with conventional tunable linear meta-devices, including all-angle beam steering and frequency-multiplexed functionalities. Our study provides new insights for realizing highly efficient and ultra-compact devices based on dynamic modulation, allowing dynamic tuning of electromagnetic waves in an almost arbitrary way, which should find potential applications in many areas including radar and compressive sensing.

\section*{Acknowledgements}
This work was supported by the Australian Research Council. The authors would like to thank Prof.~Yuri Kivshar for the valuable comment and Dr.~Momchil Minkov for the insightful discussion. 

\section*{Author Contribution}
M.~Liu conceived the idea, performed the theoretical, numerical and experimental studies, with support from  D.~A.~Powell, Y.~Zarate and I.~V.~Shadrivov in the microwave experiment.

\appendix


\section{ Boundary conditions}\label{sec:boundcon}

The physics behind the concept of time-varying Huygens' metasurfaces can be described by the generalized sheet transition conditions (GSTCs)~\cite{idemen1987boundary}: 
 \begin{eqnarray}\label{eq:fullbound1}
\mathbf{n}\times\left(\mathbf{H}_{\Vert}^{+}-\mathbf{H}_{\Vert}^{-}\right)&=&\frac{{\rm d}}{{\rm d}t}\mathbf{P_{\Vert}}-\frac{1}{\mu}\mathbf{n}\times\nabla\mathbf{M}_{\bot}\\
\mathbf{n}\times\left(\mathbf{E}_{\Vert}^{+}-\mathbf{E}_{\Vert}^{-}\right)&=&-\frac{{\rm d}}{{\rm d}t}\mathbf{M_{\Vert}}-\frac{1}{\epsilon}\mathbf{n}\times\nabla\mathbf{P}_{\bot}\label{eq:fullbound2}
\end{eqnarray}
where $\mathbf{n}$ is the normal vector of the surface; `$+$' and `$-$' represent field components on the two sides of the surface; `${\bot}$' and `${\Vert}$' denote the components being normal and parallel to the surface.  In the paper, we limit our discussion to a flat meta-device with polarizabilities only in the transverse direction ($\mathbf{P_{\bot}}=\mathbf{M_{\bot}}=0$), which is described by Eqs.~(\ref{eq:pfullbound1}) and (\ref{eq:pfullbound2}). Without loss of generality, we assume that the meta-device is on the plane  $z=0$, with $\mathbf{P}=P_x\cdot {\bf x}+P_y\cdot {\bf y}$ and $\mathbf{M}=M_x\cdot {\bf x}+M_y\cdot {\bf y}$, and the wave is propagating in the $y-z$ plane. 

When a slow periodic modulation is introduced, the time-dependent components of Eqs.~(\ref{eq:pfullbound1}) and (\ref{eq:pfullbound2}) can be decomposed in the frequency domain in the form of $F(\mathbf{r},t)=\sum_{n=-\infty}^{+\infty}F_{n}(\mathbf{r})e^{-{\rm i}\omega_{n}t}$, with $\omega_{n}=\omega_{0}+n\Omega$, with $\omega_0$ and $\Omega$ being the carrier frequency of the incident field and the modulation base frequency of the polarizations, respectively. The boundary conditions Eqs. (\ref{eq:pfullbound1}) and (\ref{eq:pfullbound2}) for TE (transverse electric) polarization at frequency $\omega_n$ are then given by
\begin{eqnarray}\label{eq:frebound1}
H_{y,n}^{\rm f}+H_{y,n}^{\rm b}-H_{y,0}^{\rm i}\delta_{n0}&=&{\rm i}\omega_{n}P_{x,n}\\
E_{x,n}^{\rm f}-E_{x,n}^{\rm b}-E_{x,0}^{\rm i}\delta_{n0}&=&{\rm i}\omega_{n}M_{y,n} \label{eq:frebound2}
\end{eqnarray}
The subscripts `i', `b' and `f' denote the incident, back-scattered and forward-scattered field components, respectively; $\delta_{n0}$ is the Kronecker delta function. The conditions for TM (transverse magnetic) polarization can be obtained from duality ($E_x \rightarrow H_x, H_y \rightarrow -E_y, P_x\rightarrow M_x, M_y\rightarrow -P_y$).

In the simple case discussed in Sec.~\ref{sec:theory_example1}, the response of the meta-device is isotropic (i.e. polarization-independent in x and y directions); the carrier wave propagates in the normal direction while the scattered sidebands are Floquet mode with transverse wave vectors $\beta_n$.  

For TE polarized propagating waves, the ratio of the transverse electric and magnetic fields is given by  $E_n/H_n=\eta k_n/\kappa_n$. $\eta=\sqrt{\mu/\epsilon}$ is the wave impedance of the surroundings, $k_n=\omega_n/c$ is the wave vector and $\kappa_n=\sqrt{k_n^2-\beta_n^2}$ is the longitudinal component. By substituting the magnetic field $H_n$ with the electric field $E_n$ in the boundary equations, we can get the transverse field components in the forward and backward directions as
\begin{eqnarray}\label{eq:scatfx}
E_{n}^{\rm f}&=&E_{n}^{\rm i}\delta_{0n}+\frac{{\rm i}\omega_{n}\eta}{2A}(\frac{k_n}{\kappa_n}p_{n}+\frac{1}{c}m_{n}),\\
E_{n}^{\rm b}&=&\frac{{\rm i}\omega_{n}\eta}{2A}(\frac{k_n}{\kappa_n}p_{n}-\frac{1}{c}m_{n}).\label{eq:scatbx}
\end{eqnarray}
$p_n=P_n A$, $m_n=M_n A/\mu$ are the Fourier components of the effective electric and magnetic dipole moments, respectively, where $A$ is the area of the unit-cell and $\mu$ is the permeability of the surroundings. 
Equations~(\ref{eq:scatfx}) and (\ref{eq:scatbx}) are employed to calculate the generalized scattering parameters: ${\rm r}_n=\sqrt{\kappa_n/k_n}E_{n}^{{\rm b}}/E^{{\rm i}}$ and ${\rm t}_n=\sqrt{\kappa_n/k_n}E_{n}^{{\rm f}}/E^{{\rm i}}$, as shown in Sec.~\ref{sec:theory_example1}. 

\section{The ideal condition for different types of sideband control}\label{sec:ideal}

Equations~(\ref{eq:scatfx}) to (\ref{eq:scatbx}) directly link the field scattered from a Huygens' meta-device to the Fourier components of the modulated electric and magnetic dipoles. By assigning the desired scattered fields $E_{n}$ in the forward and backward directions, we can retrieve the required Fourier components of the modulated dipole moments  $p_{n}$ and $m_{n}$ based on Eq.~(\ref{eq:scatfx}) and (\ref{eq:scatbx}). 

\begin{figure}[t!]
\includegraphics[width=1\columnwidth]{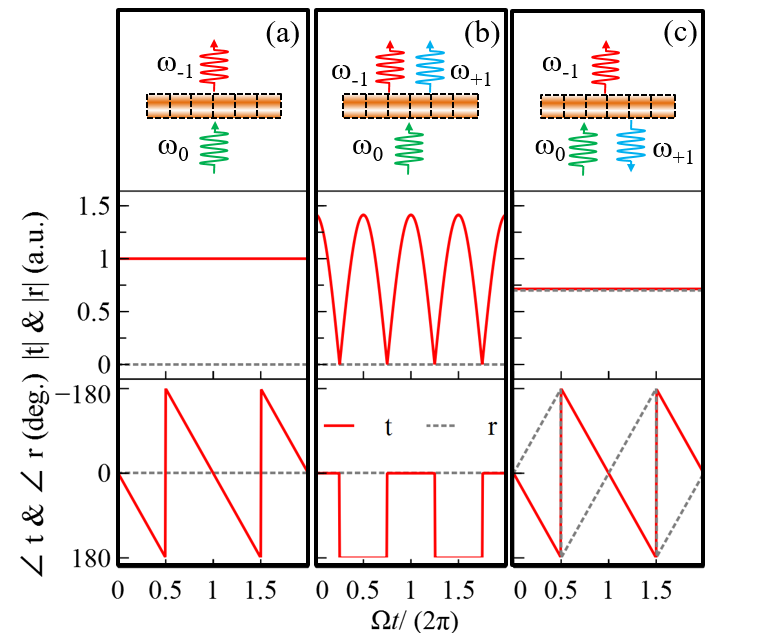}
\caption{ The normalized time-varying signals ${\rm t}=E^{\rm f }/E^{\rm i}$ and ${\rm r}=E^{\rm b }/E^{\rm i}$, for the three types of ideal directive sideband scattering: (a) single-sideband forward scattering, (b) double-sideband forward scattering, and (c) double-sideband bi-directional scattering.  \label{fig:theory_sideband} }
\end{figure}

As an example, we examine the required dipole moments for the three types of perfect directive sideband scattering that are discussed in Sec.~\ref{sec:designHU}, where all the sidebands are scattering in the normal direction ($\kappa=k$), as indicated in Fig.~\ref{fig:theory_sideband}. In the ideal situation where all the energy from the carrier wave is converted to the desired sidebands: $E_0^{\rm f}=E_0^{\rm b}=0$ and $\Sigma_{n\neq 0}|E^{\rm f}_n|^2+|E^{\rm b}_n|=|E^{\rm i}|^2$, it requires that $p_{0}=m_{0}/c=\mathrm{i}E^{\rm i}A/(\omega_0\eta)$. 

We can get the time varying electric fields $E^{\rm f(b)}(t)=\Sigma_n E^{\rm f(b)}_n  e^{-{\rm i}n\Omega t}$, as well as the required dipole moments $p_{n}$ and $m_{n}$ for the three different types of directive sideband scattering:
 
(a) single-sideband forward scattering
\begin{eqnarray} 
E^{\rm f}(t)&=&E_{-1}^{\rm f}e^{{\rm i}\Omega t}=E^{\rm i}e^{{\rm i}\Omega_0 t},\\
E^{\rm b}(t)&=&0,
\end{eqnarray}
which requires
\begin{eqnarray} 
p_{-1}=m_{-1}/c,
\end{eqnarray}
where 
\begin{eqnarray}
 |p_{-1}|=E^{\mathrm i}A/(\omega_{-1}\eta).
 \end{eqnarray}

(b) double-sideband forward scattering
\begin{eqnarray} 
E^{\rm f}(t)&=&E_{-1}^{\rm f}e^{{\rm i}\Omega t}+E_{1}^{\rm f}e^{-{\rm i}\Omega t}=\sqrt{2}E^{\rm i}\cos(\Omega t),\\
E^{\rm b}(t)&=&0,
\end{eqnarray}
which requires
\begin{eqnarray} 
p_{\pm1}=m_{\pm1}/c,
\end{eqnarray}
where 
\begin{eqnarray}
 |p_{\pm1}|=E^{\mathrm i}A/(\sqrt{2}\omega_{\pm1}\eta).
 \end{eqnarray}

(c) double-sideband bi-directional scattering
\begin{eqnarray} 
E^{\rm f}(t)&=&E_{-1}^{\rm f}e^{{\rm i}\Omega t}=E^{\rm i}e^{{\rm i}\Omega t}/\sqrt{2},\\
E^{\rm b}(t)&=&E_{1}^{\rm f}e^{-{\rm i}\Omega t}=E^{\rm i}e^{-{\rm i}\Omega t}/\sqrt{2},
\end{eqnarray}
which requires
\begin{eqnarray} 
p_{\pm1}=\mp m_{\pm1}/c,
\end{eqnarray}
where 
\begin{eqnarray}
 |p_{\pm1}|=E^{\mathrm i}A/(\sqrt{2}\omega_{\pm1}\eta).
 \end{eqnarray}

The resulting normalized time-varying signals ${\rm t}(t)=E^{\rm f }(t)/E^{\rm i}$ and ${\rm r}(t)=E^{\rm b }(t)/E^{\rm i}$ are given in  Fig.~\ref{fig:theory_sideband}. In the case where only one sideband exists on one side (front or back side) of the meta-device, it requires a pure linear phase modulation of $\rm t$ or/and $\rm r$ with a phase variation of $360^{\circ}$ [see Figs.~\ref{fig:theory_sideband} (a) and (c)]. When both sidebands exist on the same side of the meta-device [see Fig.~\ref{fig:theory_sideband} (b)], a pure sinusoidal amplitude modulation is required. 
It should be noted that there exist instances where the amplitude of the normalized signal goes above one for double-sideband forward scattering. Such an oscillation is due to the temporal interference of the two sideband waves, but the total energy remains conserved after time averaging over one modulation cycle. 

However, we do notice that in the strict adiabatic limit of modulation ($\Omega/\omega_0\rightarrow 0$), the signals $\rm t$ and $\rm r$ at each time-instance should correspond to the stationary transmission and reflection coefficients of the carrier wave, which should not exceed one at any instance  for a  purely passive meta-device. Therefore, the required signal in Fig.~\ref{fig:theory_sideband} (b) does imply that at least for adiabatic modulation, balanced gain and loss are required during each modulation cycle in order to achieve a 100\% lossless conversion to double sidebands. This is related to the recent findings in static Huygens meta-devices, which showed that balanced spatial-dependent gain and loss are required for perfect beam-steering \cite{estakhri2016wave} in the absence of bianisotropy or spatial dispersion \cite{wong2016reflectionless, PhysRevLett.117.256103}.  

\section{The interaction impedance matrix}\label{sec:impmatrix}

In general, a mutual interaction exists when the meta-atom is positioned in an array or in certain environments such as a waveguide or close to a ground-plane. The linear response of the meta-atom at position $\mathbf{r}_j$ can be described by the coupled equation:
\begin{eqnarray}
Z_{\rm self}(\mathbf{r}_j)I(\mathbf{r}_j)+\sum_{i\neq j} Z_{\rm mut}(\mathbf{r}_j,\mathbf{r}_{\rm i})I(\mathbf{r}_{\rm i}) = V_{\rm i}(\mathbf{r}_j).\nonumber\\ \label{eq:Z_inter}
\end{eqnarray}
$Z_{\rm self}(\mathbf{r}_j)$ is the self-impedance of the meta-atom at position $\mathbf{r}_j$, and $Z_{\rm mut}(\mathbf{r}_j,\mathbf{r}_{\rm i})$ is the mutual-impedance that describe the interaction between meta-atoms at $\mathbf{r}_m$ and $\mathbf{r}_n$, and $V_{\rm i}(\mathbf{r}_j)$ is the effective electromotive force (input voltage). When dynamic modulation is introduced, $Z_{\rm self}(\mathbf{r}_m)$ can be expanded into a matrix to describe the nonlinear parametric process that occurs locally within the meta-atom, while $Z_{\rm mut}(\mathbf{r}_m,\mathbf{r}_n)$ can be expanded into a matrix to describe the linear mutual interaction at each individual sideband frequency (as will be shown below).

The effect of mutual interaction can be incorporated with the self-impedance as an effective impedance $Z_{\rm eff}$, and Eq.~(\ref{eq:Z_inter}) can be simplified as 
\begin{eqnarray}
Z_{\rm eff}(\mathbf{r}_j)I(\mathbf{r}_j) = V_{\rm i}(\mathbf{r}_j). \label{eq:Z_inter2}
\end{eqnarray}
For simplicity, we have written the effective impedance of the electric (magnetic) meta-atom as $Z_{\rm E(M)}$ in the main text, and its interaction with the incident wave can be expressed as $Z_{\rm E(M)}I_{\rm E(M)}=V_{\rm E(M)}$. Note that if we retrieve the effective impedance of the meta-atom in a periodic array or waveguide via the scattering parameters from full-wave simulations, then the interaction effects are automatically taken into account.

The form of $V_{\rm i}$ depends on the incident polarization and the mode profile of the meta-atom. For simplicity, we assume that the incident wave propagates in the $y-z$ plane, the response of the meta-atoms is isotropic in the $x-y$ plane, and the meta-atoms are electrically thin in the $z$ direction. The effective electromotive force acting on the electric and magnetic meta-atoms under a plane wave excitation $\mathbf{E}^{\rm i}e^{{\rm i}(\kappa z+\beta y)}$ can be expressed explicitly. For TE wave excitation, 
\begin{eqnarray}
V_{\rm E}^{(\rm TE)}&=&\int\mathbf{j}_{\rm E}\cdot\mathbf{E}^{\rm i}{\rm d}^{3}\mathbf{r}=E^{\rm i}u_{\rm E}e^{{\rm i}\beta y}, \label{eq:VE_TE}\\
V_{\rm M}^{(\rm TE)}&=&\int\mathbf{j}_{\rm M}\cdot\mathbf{E}^{\rm i}{\rm d}^{3}\mathbf{r}={\rm i}\kappa E^{\rm i}u_{\rm M}e^{{\rm i}\beta y}.\label{eq:VM_TE}
\end{eqnarray}
For TM wave, we have
\begin{eqnarray}
V_{\rm E}^{\rm (TM)}&=&\int\mathbf{j}_{\rm E}\cdot\mathbf{E}^{\rm i}{\rm d}^{3}\mathbf{r}=E^{\rm i}u_{\rm E}e^{{\rm i}\beta y}\kappa/k, \label{eq:VE_TM}\\
V_{\rm M}^{\rm (TM)}&=&\int\mathbf{j}_{\rm M}\cdot\mathbf{E}^{\rm i}{\rm d}^{3}\mathbf{r}={\rm i}k E^{\rm i}u_{\rm M}e^{{\rm i}\beta y}.\label{eq:VM_TM},
\end{eqnarray}
where $\kappa$, $\beta$ and $k$ are the longitudinal, transverse and total wave vectors, respectively. $\mathbf{j}_{\mathrm E}(\mathbf{r})$ and $\mathbf{j}_{\mathrm M}(\mathbf{r})$ are the normalized electric current distributions of the modes on the electric and magnetic meta-atoms. At normal incidence, the expressions are simplified as Eqs.~(\ref{eq:VE}) and (\ref{eq:VM}), respectively. $u_{\rm E}$ and $u_{\rm M}$ are the normalized effective electric and magnetic dipole moments of the meta-atoms, which can be defined from the normalized electric current distribution $\bf j (\mathbf{r})$ 
\begin{eqnarray}
u_{\rm E}&=&\int\mathbf{j}_{\rm E}(\mathbf{r}){\rm d}^{3}\mathbf{r} \label{eq:Edipolemoment},\\
u_{\rm M}&=&\frac{1}{2}\int\mathbf{r} \times\mathbf{j}_{\rm M}(\mathbf{r}){\rm d}^{3}\mathbf{r}, \label{eq:Mdipolemoment} 
\end{eqnarray}
where the integral is performed over the volume of the meta-atom. Note that since the normalized  current $\bf j (\mathbf{r})$ has a unit of $\rm m^{-2}$, $u_{\rm E}$ has a unit of $\rm m$, while $u_{\rm M}$ has a unit of $\rm m^2$. 

The effective impedance can also be defined rigorously based on the mode of the meta-atoms, where the details can be found in our previous studies~\cite{liu2012optical,PhysRevB.90.075108}. \\

 \begin{figure}[t!]
\includegraphics[width=0.5\columnwidth]{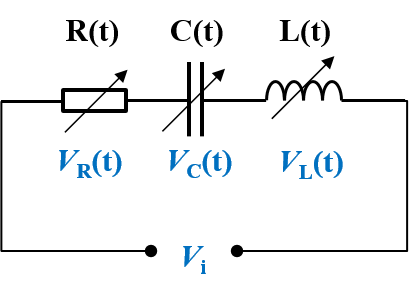}
\caption{ Schematic of the series RLC equivalent circuit for a time-varying meta-atom.  \label{fig:RLC} }
\end{figure}

$\,$\\

Below, we derive Eq.~(\ref{eq:impedanceFourier}) in the main text.
When dynamic modulation is introduced to the meta-atom at position $\mathbf{r}_j$, the effective impedance $Z_{\rm eff}$, mode amplitude $I$ and effective electromotive force $V_{\rm i}$ become time-dependent. Based on Kirchoff's voltage law, the input voltage should equal the total voltage across the series RLC equivalent circuit (see Fig.~\ref{fig:RLC}):
\begin{eqnarray}\label{eq:kirch}
V_{\rm i}(t)	=	V_{\rm L}(t)+V_{\rm C}(t)+V_{\rm R}(t)
\end{eqnarray}
where
\begin{eqnarray}
V_{\rm L}(t)&=&\frac{\mathrm{d}LI}{\mathrm{d}t}=I(t)\frac{\mathrm{d}L}{\mathrm{d}t}+L(t)\frac{\mathrm{d}I}{\mathrm{d}t}\nonumber\\&=&\dot{Q}(t)\dot{L}(t)+L(t)\ddot{Q}(t),\\
V_{\rm C}(t)&=&\frac{Q(t)}{C(t)}=S(t)Q(t),\\
V_{\rm R}(t)&=&R(t)\dot{Q}(t)
\end{eqnarray}
Here $S=1/C$ is the elastance, and we use charge $Q$ since it yields purely differential equations rather than integro-differential equations. $V_{\rm i}$ is the input voltage generated by a particular set of polarization and mode, represented by one of  Eqs.~(\ref{eq:VE_TE}) to (\ref{eq:VM_TM}). For illustration, we use $V_{\rm i}(t)={E}^{\rm i}{u}_{E}(t).$ 

When the temporal modulation of the impedance is periodic in time with a frequency $\Omega \ll\omega$, each time varying effective circuit parameter in the above equations can be expanded in a Fourier series of the form: $F(t)=\sum_{l=-\infty}^{+\infty}F_{l}e^{{\rm -i}l(\Omega t-\varphi)}$, and the charge can also be expanded as $Q(t)=e^{-{\rm i}\omega t}\sum_{l=-\infty}^{+\infty}Q_{l}e^{{\rm -i}l(\Omega t-\varphi)}=\sum_{l=-\infty}^{+\infty}Q_{l}e^{-{\rm i}\omega_{l}t}e^{{\rm i}l\varphi}$. $\varphi$ is the modulation phase corresponding to a constant time-offset for the modulation waveform. Then we have
\begin{widetext}
\begin{eqnarray}
V_{\rm L}(t)&=&\left[-{\rm i}\sum_{m=-\infty}^{+\infty}\omega_{m}Q_{m}e^{-{\rm i}\omega_{m}t}e^{{\rm i}m\varphi}\right]\times\left[-{\rm i}\sum_{l=-\infty}^{+\infty}l\Omega L_{l}e^{-{\rm i}l\Omega t}e^{{\rm i}l\varphi}\right]\\\nonumber&+&\left[-\sum_{m=-\infty}^{+\infty}\omega_{m}^{2}Q_{m}e^{-{\rm i}\omega_{m}t}e^{{\rm i}m\varphi}\right]\times\left[\sum_{l=-\infty}^{+\infty}L_{l}e^{-{\rm i}l\Omega t}e^{{\rm i}l\varphi}\right]\\
V_{\rm C}(t)&=&\left[\sum_{l=-\infty}^{+\infty}S_{l}e^{-{\rm i}l\Omega t}e^{{\rm i}l\varphi}\right]\times\left[\sum_{m=-\infty}^{+\infty}Q_{m}e^{-{\rm i}\omega_{m}t}e^{{\rm i}m\varphi}\right],\\
V_{\rm R}(t)&=&\left[\sum_{l=-\infty}^{+\infty}R_{l}e^{-{\rm i}l\Omega t}e^{{\rm i}l\varphi}\right]\times\left[-{\rm i}\sum_{m=-\infty}^{+\infty}\omega_{m}Q_{m}e^{-{\rm i}\omega_{m}t}e^{{\rm i}m\varphi}\right].
\end{eqnarray}
For electric meta-atoms excited by a normally incident field
\begin{eqnarray}
V_{\rm i}(t)&=&E^{\rm i}\sum_{n=-\infty}^{+\infty}u_{{\rm E},n} e^{-{\rm i}\omega_{n}t}e^{{\rm i}n\varphi}.
\end{eqnarray}

Satisfying Eq.~(\ref{eq:kirch}) at every instant in time requires that the Fourier coefficients satisfy the following equation

\begin{eqnarray}\label{eq:coff0}
\sum_{l=-\infty}^{+\infty}\sum_{m=-\infty}^{+\infty}\left[-(\omega_{m}l\Omega+\omega_{m}^{2})L_{l}+S_{l}-{\rm i}\omega_{m}R_{l}\right]Q_{m}e^{-{\rm i}(l+m)(\Omega t-\varphi)}&=&E^{\rm i}\sum_{n=-\infty}^{+\infty}u_{{\rm E},n}e^{{\rm -i}n(\Omega t-\varphi)}.
\end{eqnarray}
Orthogonality of the exponential functions implies that equivalent terms on each side of Eq.~\eqref{eq:coff0} must be equal. This yields the following equation for each order $n$:

\begin{eqnarray}\label{eq:coff1}
\sum_{m=-\infty}^{+\infty}\left[-(\omega_{m}l\Omega+\omega_{m}^{2})L^{(l)}+S^{(l)}-{\rm i}\omega_{m}R^{(l)}\right]Q^{(m)}e^{-{\rm i}(l+m)(\Omega t-\varphi)}&=&E^{\rm i}u_{{\rm E}, n}e^{{\rm -i}n(\Omega t-\varphi)}.
\end{eqnarray}
where  $l=n-m$. We further define $I_m=-i\omega_m Q_m$, $Z_{l}^{(m)}=-i(\omega_{m}+l\Omega)L_l+iS_l/\omega_{m}+R_l$, and $V_n=E^{i}u_{{\rm E},n}$, then Eq.~\eqref{eq:coff1} becomes 
\begin{eqnarray}\label{eq:coff2}
\sum_{m=-\infty}^{+\infty}Z_l^{(m)}I_{m}e^{-{\rm i}(l+m)(\Omega t-\varphi)}=V_{n}e^{-{\rm i}n(\Omega t-\varphi)},\nonumber\\
\end{eqnarray}
which is Eq.~(\ref{eq:impedanceFourier}) in the main text. Note that for brevity, we have omitted the notation $\mathbf{r}_j$ in the above equations.

Now we bring the position notation $\mathbf{r}_j$ back and write the coefficients of Eq.~(\ref{eq:coff2}) in a matrix form:
\begin{eqnarray}\label{eq:impedancematrix_A}
\overleftrightarrow{\mathbf{Z}_{\rm eff}}(\mathbf{r}_j)\mathbf{I}(\mathbf{r}_j)&=&\mathbf{V}(\mathbf{r}_j)
\end{eqnarray}
where

\begin{eqnarray}\label{eq:impedance_eff}
\overleftrightarrow{\mathbf{Z}_{\rm eff}}(\mathbf{r}_j)=&\left[\begin{array}{ccccc}
\ddots & \vdots &  & \vdots\\
\cdots & Z_{\rm 0}^{(m)}(\mathbf{r}_j) & \cdots & Z_{m-n}^{(n)}(\mathbf{r}_j) e^{{\rm i}(m-n)\varphi_j} &\cdots\\
 & \vdots & \ddots & \vdots\\
\cdots & Z_{n-m}^{(m)}(\mathbf{r}_j)e^{{\rm i}(n-m)\varphi_j } & \cdots & Z_{\rm 0}^{(n)}(\mathbf{r}_j) & \cdots\\
 & \vdots &  & \vdots & \ddots
\end{array}\right],
\end{eqnarray}
\begin{eqnarray}\label{eq:modeamp}
\mathbf{I}(\mathbf{r}_j)=\left[\begin{array}{c}
\cdots, I_{m}(\mathbf{r}_j)e^{{\rm i}m\varphi_j},\cdots,I_{n}(\mathbf{r}_j)e^{{\rm i}n\varphi_j},\cdots
\end{array}\right]^T,
\end{eqnarray}
\begin{eqnarray}
\mathbf{V}(\mathbf{r}_j)=\left[\begin{array}{c}
\cdots, V_{m}(\mathbf{r}_j)e^{{\rm i}m\varphi_j},\cdots, V_{n}(\mathbf{r}_j)e^{{\rm i}n\varphi_j},\cdots
\end{array}\right]^T.
\end{eqnarray}
\end{widetext}
$\varphi_j$ is the modulation phase at $\mathbf{r}_j$. It is important to note that while we have incorporated the effect of mutual interaction in the effective impedance $\overleftrightarrow{\mathbf{Z}_{\rm eff}}$, we have assumed that only the modulation of self-impedance contributes to the higher order terms $Z_l^{(m)} (l\neq 0)$, i.e. 
\begin{eqnarray}\label{eq:impedance_SM}
\overleftrightarrow{\mathbf{Z}_{\rm eff}}(\mathbf{r}_j)\mathbf{I}(\mathbf{r}_j)&=&\overleftrightarrow{\mathbf{Z}_{\rm self}}(\mathbf{r}_j)\mathbf{I}(\mathbf{r}_j)+\sum_i\overleftrightarrow{\mathbf{Z}_{\rm mut}}(\mathbf{r}_j,\mathbf{r}_i)\mathbf{I}(\mathbf{r}_i)\nonumber\\
\end{eqnarray} 
where 
\begin{widetext}
\begin{eqnarray}\label{eq:impedance_S}
\overleftrightarrow{\mathbf{Z}_{\rm self}}(\mathbf{r}_j)=&\left[\begin{array}{ccccc}
\ddots & \vdots &  & \vdots\\
\cdots & Z_{\rm self}^{(m)}(\mathbf{r}_j) & \cdots & Z_{m-n}^{(n)}(\mathbf{r}_j) e^{{\rm i}(m-n)\varphi_j} &\cdots\\
 & \vdots & \ddots & \vdots\\
\cdots & Z_{n-m}^{(m)}(\mathbf{r}_j)e^{{\rm i}(n-m)\varphi_j } & \cdots & Z_{\rm self}^{(n)}(\mathbf{r}_j) & \cdots\\
 & \vdots &  & \vdots & \ddots
\end{array}\right],
\end{eqnarray}
 \end{widetext}

$Z_{\rm self}^{(m)}(\mathbf{r}_j)$ is the zeroth order self-impedance of the meta-atom at position $\mathbf{r}_j$, evaluated at frequency $\omega_m$. The mutual impedance matrix describes the linear interaction at each sideband frequency, and thus it is a diagonal matrix with only the zeroth order terms 
\begin{eqnarray}\label{eq:impedance_M}
\overleftrightarrow{\mathbf{Z}_{\rm mut}}(\mathbf{r}_j,\mathbf{r}_i)=&\left[\begin{array}{ccccc}
\ddots & \vdots &  & \vdots\\
\cdots & Z_{\rm mut}^{(m)}(\mathbf{r}_j,\mathbf{r}_i) & \cdots & 0 &\cdots\\
 & \vdots & \ddots & \vdots\\
\cdots & 0 & \cdots & Z_{\rm mut}^{(n)}(\mathbf{r}_j,\mathbf{r}_i) & \cdots\\
 & \vdots &  & \vdots & \ddots
\end{array}\right]. \nonumber\\
\end{eqnarray}
Therefore, the value of the zeroth order effective impedance $Z_0^{m}$ in Eq.~(\ref{eq:impedance_eff}) is determined by not only the zeroth order self-impedance but also the mutual impedances and the mode amplitudes of all other interacting meta-atoms at sideband frequency $\omega_m$. In general, $Z_0^{(m)}\neq Z_0^{(n)}$ if $m\neq n$, except for some special cases.

When the modulation frequency $\Omega$ is much smaller than the linewidth of resonance of the meta-atoms, the difference between the impedance elements of the same order evaluated at different sideband frequencies $\omega_l$ and $\omega_m$ becomes negligible. We can simplify the notation as $Z_{\rm self}=Z_{\rm self}^{(m)}=Z_{\rm self}^{(l)}$,  $Z_{ n}=Z_{ n}^{(m)}=Z_{ n}^{(l)}$ and $Z_{\rm mut}=Z_{\rm mut}^{(m)}=Z_{\rm mut}^{(l)}$, and the impedance matrices Eq.~(\ref{eq:impedance_S}) and (\ref{eq:impedance_M}) can be simplified as Toeplitz matrices:
\begin{eqnarray}\label{eq:impedance_SToep}
\overleftrightarrow{\mathbf{Z}_{\rm self}}(\mathbf{r}_j)=\left[\begin{array}{ccccc}
\ddots & \vdots &  & \vdots\\
\cdots & Z_{\rm self}(\mathbf{r}_j) & \cdots & Z_{n}(\mathbf{r}_j)  e^{{\rm i}n\varphi_j} &\cdots\\
 & \vdots & \ddots & \vdots\\
\cdots & Z_{-n}(\mathbf{r}_j)e^{-{\rm i}n\varphi_j } & \cdots & Z_{\rm self}(\mathbf{r}_j) & \cdots\\
 & \vdots &  & \vdots & \ddots
\end{array}\right]\nonumber\\\\
\overleftrightarrow{\mathbf{Z}_{\rm mut}}(\mathbf{r}_j,\mathbf{r}_i)=\left[\begin{array}{ccccc}
\ddots & \vdots &  & \vdots\\
\cdots & Z_{\rm mut}(\mathbf{r}_j,\mathbf{r}_i) & \cdots & 0 &\cdots\\
 & \vdots & \ddots & \vdots\\
\cdots & 0 & \cdots & Z_{\rm mut}(\mathbf{r}_j,\mathbf{r}_i) & \cdots\\
 & \vdots &  & \vdots & \ddots
\end{array}\right]. \label{eq:impedance_MToep}\nonumber\\
\end{eqnarray}

For the special example discussed in Sec.~\ref{sec:theory_example1}, where the dynamic modulation of meta-atoms has an identical modulation amplitude but a periodic linear phase gradient along $y$ direction: $\varphi(y)=\varphi_0+Gy$, the solution of the system Eq.~(\ref{eq:modeamp}) should be a series of Floquet modes:
\begin{eqnarray}
\mathbf{I}(y_j)=\mathbf{I}(y_i)\overleftrightarrow{{\bf G}}(y_j, y_i) 
\end{eqnarray} 
where 
\begin{eqnarray}
\mathbf{I}(y_j)&=&\left[\begin{array}{c}
\cdots, I_{n}(y_j)e^{{\rm i}(n\varphi_0+\beta_n y_j)},\cdots
\end{array}\right]^T,\nonumber\\\\
\overleftrightarrow{{\bf G}}(y_j, y_i) &=& \left[\begin{array}{ccccc}
\ddots & \vdots &  & \vdots\\
\cdots & e^{-\mathrm{i}nG(y_j-y_i)} & \cdots & 0 &\cdots\\
 & \vdots & \ddots & \vdots\\
\cdots & 0 & \cdots &  e^{\mathrm{i}nG(y_j-y_i)} & \cdots\\
 & \vdots &  & \vdots & \ddots
\end{array}\right].\label{eq:Gmatrix} \nonumber\\
\end{eqnarray} 
$G$ is the spatial frequency of modulation, $\beta_n=\beta_0+nG$ is the transverse wave vector of the Floquet mode, and $\varphi_0$ is the modulation phase at $y=0$. 

Applying Eqs.~(\ref{eq:impedance_SToep}) to (\ref{eq:Gmatrix}) in Eq.~(\ref{eq:impedance_SM}), it becomes clear that the effective impedance matrix of the series of Floquet modes is given by
\begin{eqnarray}\label{eq:impedance_Floq}
\overleftrightarrow{\mathbf{Z}_{\rm eff}}(y_j)=\overleftrightarrow{\mathbf{Z}_{\rm self}}(y_j)+\sum_i\overleftrightarrow{\mathbf{Z}_{\rm mut}}(y_j,y_i)\overleftrightarrow{{\bf G}}(y_j, y_i)\nonumber\\=\left[\begin{array}{ccccc}
\ddots & \vdots &  & \vdots\\
\cdots & Z_{\beta,m} & \cdots & Z_{m-n}  e^{{\rm i}(m-n)\varphi_j} &\cdots\\
 & \vdots & \ddots & \vdots\\
\cdots & Z_{n-m}e^{-{\rm i}(n-m)\varphi_j } & \cdots & Z_{\beta,n} & \cdots\\
 & \vdots &  & \vdots & \ddots
\end{array}\right]\nonumber\\
\end{eqnarray} 
where the higher order components $Z_{n}$ are generated directly from the modulation of self-impedance, while $Z_{\beta,m}(y_j)=Z_{\rm self}(y_j)+\sum_i Z_{\rm mut}(y_j, y_i)e^{\mathrm{i}mG(y_j-y_i)}$ is the zeroth order effective impedance of Floquet mode $(\omega_m, \beta_m)$.  With a normally incident carrier wave ($\beta_0=0$), we have $Z_{\beta,m}=Z_{\beta,-m}$, and the effective impedance matrix is given by Eq.~(\ref{eq:imp_ntoeplitz}). Only in the special case where the sidebands are also scattered in the normal direction ($G=0$), the effective impedance matrix becomes a Toeplitz matrix:
\begin{eqnarray} \label{eq:imp_toeplitz}
\overleftrightarrow{\mathbf{Z}}&=&\left[\begin{array}{ccccc}
Z_{0} & \cdots & Z_{-n}e^{-{\rm i}n\varphi} & \cdots & Z_{-2n}e^{-{\rm i}2n\varphi}\\
\vdots & \ddots & \vdots & \ddots & \vdots\\
Z_{n}e^{{\rm i}n\varphi} & \cdots & Z_{0} & \cdots & Z_{-n}e^{-{\rm i}n\varphi}\\
\vdots & \ddots & \vdots & \ddots & \vdots\\
Z_{2n}e^{{\rm i}2n\varphi} & \cdots & Z_{n}e^{{\rm i}n\varphi} & \cdots & Z_{0}
\end{array}\right]\nonumber\\
\end{eqnarray}

Generally, the Fourier coefficient $Z_n$ and $Z_{-n}$ are complex but not conjugated. In the special situation where only reactive modulation exists, $Z_1=-Z_{-1}^{*}=-\rm{i}\xi$, and the phase of the complex value $\xi$ is determined by the time delay of the modulation waveform. Since we already introduce the phase parameter $\varphi$ to describe the time delay of the modulation signal, it is convenient to set the phase of $\xi$ to zero in Eq.~(\ref{eq:impFloq2}) such that $\xi\in\mathbb{R}$.   

\section{The relation between radiative loss and normalized dipole moments}\label{sec:radloss}

Below we derive the relation between the radiative loss term $R^{(\rm rad)}$ and the normalized dipole moments for the Floquet mode ($\omega_n, \beta_n$) under TE polarization, as shown in Eq.~(\ref{eq:radloss}) of the main text:
\begin{eqnarray}
R_{{\rm E},n}^{(\rm rad)}=\frac{\eta u_{\rm E}^{2}k_n}{2A\kappa_n},\,\, R_{{\rm M},n}^{(\rm rad)}=\frac{\eta u_{\rm M}^{2}k_n\kappa_n}{2A}.
\end{eqnarray}
Note that the above relations are only valid for Floquet mode ($\omega_n, \beta_n$) under TE polarization. These intrinsic relations are determined by the current distributions of the modes, and thus should hold regardless of the modulation condition. Therefore, it is convenient to derive the relations in an unmodulated meta-device consisting of identical meta-atoms in each unit, where both the excitation and the scattering are propagating plane waves with a transverse wave vector $\beta_n$.  For a lossless meta-device ($R^{(\rm ohm)}=0$), the fields in the forward and backward directions are given by
 \begin{eqnarray}\label{eq:scatf}
E^{{\rm f}}&=&E^{{\rm i}}+E^{\rm p}+E^{\rm m},\\\label{eq:scatb}
E^{{\rm b}}&=&E^{\rm p}-E^{\rm m}.
\end{eqnarray}
where $E^{\rm p}$ and $E^{\rm m}$ are the scattered plane waves generated by the array of electric and magnetic meta-atoms, respectively. From the passivity condition, the power should be conserved: 
 \begin{eqnarray}\label{eq:powercons}
|E^{{\rm f}}|^2+|E^{{\rm b}}|^2&=&|E^{{\rm i}}|^2, 
\end{eqnarray}
From Eqs.~(\ref{eq:scatf}), (\ref{eq:scatb}) and (\ref{eq:powercons}), we have the following relation:
\begin{eqnarray}\label{eq:powercons1}
{\rm Re}[(E^{\rm i})^*E^{\rm p}+(E^{\rm i})^*E^{\rm m}]+|E^{\rm p}|^2+|E^{\rm m}|^2=0.
\end{eqnarray}
This relation should hold regardless of the ratio of $E^{\rm p}$ and $E^{\rm m}$, i.e. it should also be satisfied even for an array of pure electric or pure magnetic meta-atoms, and thus requires the electric fields generated by electric and magnetic meta-atoms satisfy the following relations individually:
 \begin{eqnarray}\label{eq:powercons2}
{\rm Re}[(E^{\rm i})^*E^{\rm p}]+|E^{\rm p}|^2&=&0,\\
{\rm Re}[(E^{\rm i})^*E^{\rm m}]+|E^{\rm m}|^2&=&0,
\end{eqnarray}
Substitute with the following relations:
 \begin{eqnarray}\label{eq:Edipolefields}
E^{\rm p}&=&\frac{{\rm i}\omega\eta k_n}{2A\kappa_n}p =\frac{{\rm i}\omega\eta k_n}{2A\kappa_n}\frac{{\rm i}E^{\rm i}u_{\rm E}^2}{\omega Z_{{\mathrm E}, \beta,n}}=-\frac{\eta u_{{\rm E}}^{2}k_n}{2A\kappa_n}\frac{E^{{\rm i}}}{Z_{\mathrm{E},\beta,n}},\nonumber\\\\
E^{\rm m}&=&\frac{{\rm i}\omega\eta}{2A}\frac{1}{c}m=\frac{{\rm i}\omega_n\eta}{2Ac}\frac{{\rm i}\kappa_n E^{{\rm i}}u_{{\rm M}}^{2}}{Z_{\mathrm{M}, \beta,n}}=-\frac{\eta u_{{\rm M}}^{2}k_n^2 }{2A}\frac{E^{{\rm i}}}{Z_{\mathrm{M},\beta,n}},\nonumber\\ \label{eq:Mdipolefields}
\end{eqnarray}
and use the identity for lossless meta-atoms ${\rm Re}(1/Z)={\rm Re}(Z^*)/|Z|^2=R^{(\rm rad)}/|Z|^2$, we get Eq.~(\ref{eq:radloss}).

\section{Evaluating the accuracy of the impedance model}\label{sec:imp}

Below, we evaluate the accuracy of the impedance model based on the Toeplitz matrix shown by Eq.~(\ref{eq:imp_toeplitz}), which is a special case of Eq.~(\ref{eq:imp_ntoeplitz}). One way to calculate the impedance matrix  Eq.~(\ref{eq:imp_toeplitz}) is to retrieve the time-varying impedance $Z(t)$ from the time-varying transmission and reflection coefficients, and calculate the Fourier coefficients $Z_n$ via a Fourier transformation of  $Z(t)$.

As can be inferred from Eqs.~(\ref{eq:r0_f}) and (\ref{eq:t0_f}), the stationary transmission and reflection coefficients of a static meta-device excited by a normally incident plane wave can be expressed as 
\begin{eqnarray} \label{eq:rs_f}
\mathrm{r}&=&-\frac{R_{{\rm E}}^{({\rm rad})}}{Z_{{\rm E}}}+\frac{R_{{\rm M}}^{({\rm rad})}}{Z_{{\rm M}}},\\
\mathrm{t}&=&1-\frac{R_{{\rm E}}^{({\rm rad})}}{Z_{{\rm E}}}-\frac{R_{{\rm M}}^{({\rm rad})}}{Z_{{\rm M}}},\label{eq:ts_f}
\end{eqnarray}
from which we can retrieve the effective impedance $Z_{{\rm E}}$ and $Z_{{\rm M}}$ as
\begin{eqnarray} \label{eq:imp_E}
Z_{{\rm E}}&=&2R_{{\rm E}}^{\rm (rad)}/(1-\mathrm{t}-\mathrm{r}),\\
Z_{{\rm M}}&=&2R_{{\rm M}}^{\rm (rad)}/(1-\mathrm{t}+\mathrm{r}) \label{eq:imp_M}
\end{eqnarray}

The time-varying impedance $Z_{{\rm E}}(t)$ and $Z_{{\rm M}}(t)$ under a slow modulation ($\tilde{U}_{{\rm E}}(t)$ and $\tilde{U}_{{\rm M}}(t)$) can then be approximated by calculating the stationary response at each time-step with Eqs.~(\ref{eq:imp_E}) and  (\ref{eq:imp_M}), and the Fourier coefficients $Z_{{\rm E (M)},n}$ can be obtained via $Z_{{\rm E (M)},n}=\frac{1}{2\pi}\int_{0}^{T} Z_{{\rm E (M)}}(t) e^{\mathrm{i}
n\Omega t}dt$, where $T=2\pi/\Omega$ is the modulation period.

Once we get the impedance matrix $\overleftrightarrow{\mathbf{Z}}_{\rm E(M)}$, we can calculate the mode amplitude $\mathbf{I}_{\rm E(M)}$ using Eq.~(\ref{eq:intmatrix}), and the corresponding dipole moments $p_n$ and $m_n$ from Eqs.~(\ref{eq:edipole}) and (\ref{eq:mdipole}), as well as the scattered fields $E^{\rm f(b)}_n$ from Eqs.~(\ref{eq:scatfx}) and (\ref{eq:scatbx}).  At normal incidence, the generalized scattering parameters can be defined as ${\rm r}_n=E_{n}^{{\rm b}}/E^{{\rm i}}$ and ${\rm t}_n=E_{n}^{{\rm f}}/E^{{\rm i}}$. The final expression of the generalized scattering parameters can be expressed in a compact form:
\begin{eqnarray} \label{eq:rs_fsb}
\mathbf{r}&=&-\frac{\mathcal{D}R_{{\rm E}}^{({\rm rad})}}{\overleftrightarrow{\mathbf{Z}}_{\rm E}}+\frac{\mathcal{D}R_{{\rm M}}^{({\rm rad})}}{\overleftrightarrow{\mathbf{Z}}_{\rm M}},\\
\mathbf{t}&=&\mathcal{D}-\frac{\mathcal{D}R_{{\rm E}}^{({\rm rad})}}{\overleftrightarrow{\mathbf{Z}}_{\rm E}}-\frac{\mathcal{D}R_{{\rm M}}^{({\rm rad})}}{\overleftrightarrow{\mathbf{Z}}_{\rm M}},\label{eq:ts_fsb}
\end{eqnarray}
where $\mathcal{D}=[\delta_{0-N},\cdots,\delta_{00},\cdots,\delta_{0N}]^{T}$, with $\delta_{0N}$ being the Kronecker delta function; ${\bf r}$=$[r_{-N},\cdots,r_{0},\cdots,r_{N}]^{T}$ and ${\bf t}$=$[t_{-N},\cdots,t_{0},\cdots,t_{N}]^{T}$. $N$ is the order of truncation.

 \begin{figure}[t!]
\includegraphics[width=1\columnwidth]{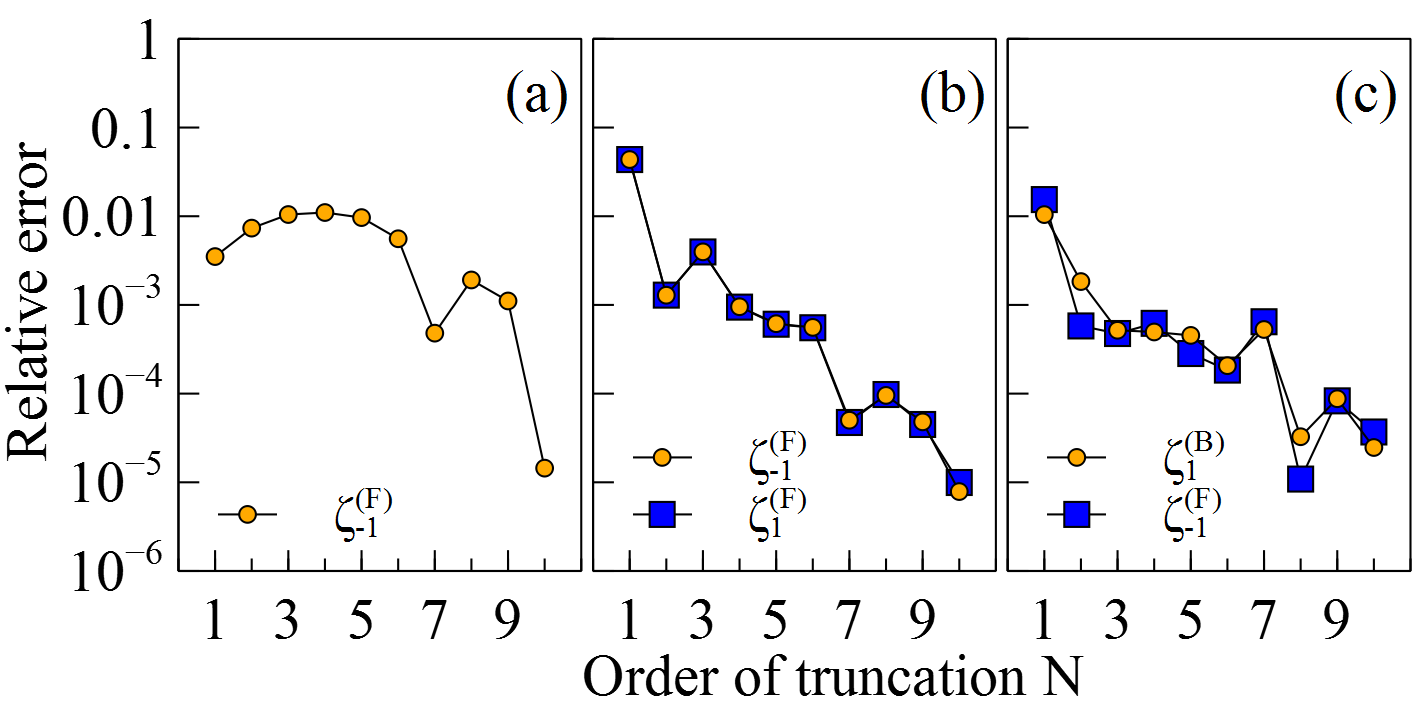}
\caption{(a) to (c)~The relative error of the sideband power calculated using the impedance model when comparing to the ones shown in Fig.~\ref{fig:numerical} (a) to (c).  \label{fig:imp_compare} }
\end{figure} 

To show the accuracy of the impedance model as a function of the order of truncation $N$, we utilize the impedance model to calculate the sideband spectra under different optimized modulation waveforms shown in Figs.~\ref{fig:numerical} (d) to (f), and evaluate the accuracy by comparing the full-wave results depicted in Figs.~\ref{fig:numerical} (a) to (c). Figures~\ref{fig:imp_compare} (a) to (c) depict the relative calculation error of the impedance model when comparing to the sideband spectra calculated with full-wave simulation:  $|\zeta_{\rm full}-\zeta_{\rm imp}|/\zeta_{\rm full}$; for clarity, we only show the relative error of the dominant sidebands in each type of modulation. It is clear that as the order of truncation increases in the impedance model, the result converges and approach to the full-wave calculation. Surprisingly, the maximum relative error in the three cases studied is already below 5\% even when we truncate to the first order sideband, as shown in Fig.~\ref{fig:imp_compare} (b) for the case of double-sideband forward scattering. Note that different from the simplified case discussed in Eq.~(\ref{eq:impFloq2}),  the second order terms $Z_{\pm 2}$ are also preserved in the impedance matrix Eq.~(\ref{eq:imp_toeplitz}) when we truncate the sidebands to $N=1$. The introduction of higher order impedance terms  $Z_n (|n|>2)$ allows us to calculate the case where the modulation waveform is asymmetric, as is required for the case of single-sideband conversion. 

\section{Numerical simulation}\label{sec:num_sim}

To validate the practicality of the concept, we design electric and magnetic meta-atoms working in the microwave regime using full wave simulation (CST Microwave studio). 

For a single Huygens' unit in a rectangular waveguide, we employ electric boundary condition on the waveguide sidewalls and open boundary condition otherwise. The meta-atoms are positioned in the middle of the waveguide and excited with the fundamental mode of the rectangular waveguide.

\begin{figure*}[t!]
\includegraphics[width=1.5\columnwidth]{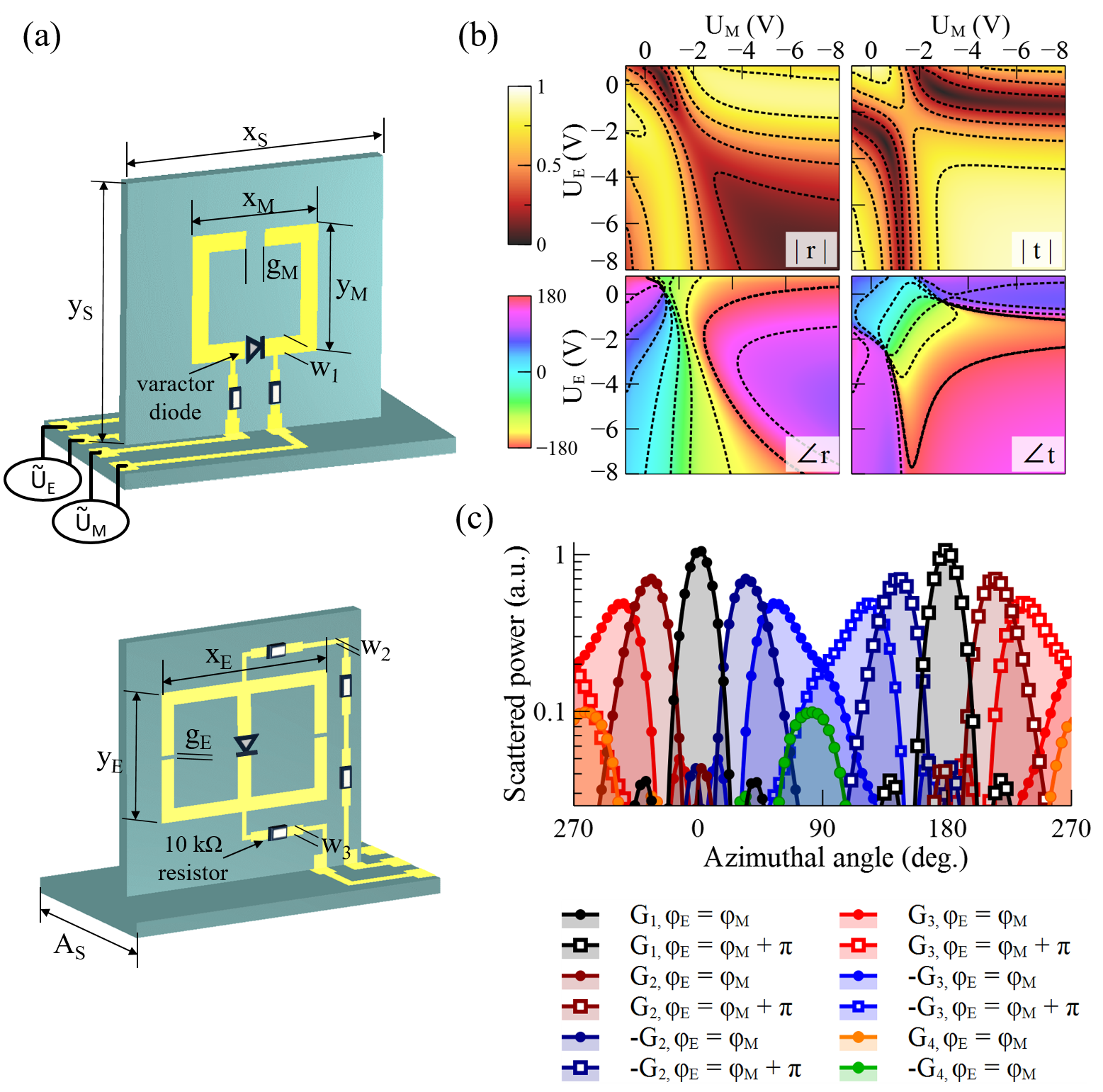}
\caption{(a) Design of the Huygens' unit of the array structure. The electric and magnetic meta-atoms are on the opposite sides of the substrate (0.8 mm thick Rogers4003C). $x_{\mathrm S}=13.52, y_{\mathrm S}=13.20, w_{1}=1.00,w_{2}=0.25, w_{3}=0.50, x_{\mathrm M}=6.64, y_{\mathrm M}=6.46, x_{\mathrm E}=9.52, y_{\mathrm E}=6.80, g_{\mathrm M}=1.00, g_{\mathrm E}=0.20, A_{\mathrm S}$=10, (mm). Operating at 4 GHz ($\lambda\approx 75$mm), the overall size of the unit ($x_{\mathrm S}\times y_{\mathrm S} \times A_{\mathrm S}$) is 0.19$\lambda \times$ 0.18$\lambda \times$ 0.13$\lambda$.  The gap between the parallel plates is around 15.5 mm ($\sim 0.21\lambda$). (b) The simulated transmission and reflection coefficients of the periodic array at 4 GHz as a function of the DC bias voltages, in the absence of modulation. (c) Simulated scattering patterns of a finite array of 16 Huygens' units.\label{fig:array_unit} }
\end{figure*}

For the Huygens' array, we re-optimize the designs in the parallel-plate waveguide. We simulate the structure under periodic boundary conditions along the direction of the array and perfect electric boundaries on the top and bottom waveguide walls. The unit is excited with the fundamental TEM mode. The detailed geometries can be found in Fig.~\ref{fig:array_unit}. To enhance the mechanical stability, the electric and magnetic meta-atoms are designed on the opposite sides of a 0.8 mm thick substrate (Rogers RO4003), and the center-to-center distance between neighboring units is 10 mm. High-resistance (10 k$\omega$) resistors are employed for the bias-lines to avoid the out-coupling of microwave signals to the external modulators.

To calculate the sidebands in the adiabatic limit, we first run a full-wave simulation of the system by replacing the diodes with discrete ports; the full impedance matrix of the system can be extracted and used to perform a circuit simulation, where we define the property of the varactor diode based on its SPICE model. The transmission and reflection coefficients under different bias voltage $U_{\rm E}$ and $U_{\rm M}$ are calculated by a parameter scan in the circuit simulation of CST. The simulated response of the array [Fig.~\ref{fig:array_unit} (b)] has the same feature as the one of a single unit in the waveguide [Fig.~\ref{fig:numerical_design} (b)]. The time-dependent transmission and reflection under periodic modulation of $U_{\rm E}(t)$ and $U_{\rm M}(t)$ are found via a one-to-one voltage-scattering parameter mapping; finally, a Fourier transformation of the time-dependent signals gives the information of sidebands in both forward and backward directions.

This adiabatic approximation is valid when the modulation frequency $\Omega$, the range of resonant frequency modulation $\Delta$ and the linewidth of the resonance $\gamma$ satisfy $\Delta\Omega\ll\gamma^2$ \cite{Momchil2017Exact}. In our studied system, $\Omega/\gamma\approx 0.009$, $\Delta/\gamma< 1.7$, therefore the adiabatic approximation is valid. This approximation could provide more physical insight and higher efficiency during optimization compared to first-principle calculation methods such as FDTD, which become very inefficient when critical time scales vary by several orders of magnitude. 

The SPICE model of the SMV1405 varactor diodes is obtained from the datasheet \cite{smv1405}, and has the following parameters: is=1e-014, rs=0.8, n=1,  eg=1.11, xti=3, ibv=1e-03, cjo=2.37e-12, cp=0.29e-12, ls=0.7e-9, vj=0.77, m=0.5, fc=0.5, tt=0, kf=0, af=1.

For the scattering of a finite array under different modulation phase patterns, due to the limited numerical simulation capacity, we can only perform a linear simulation to give a reference, as we assume that the mode profiles of the meta-atoms will not have a substantial difference at the sidebands and the carrier frequency when the modulation frequency satisfies the adiabatic approximation. To do so, we perform a linear simulation of a periodic array and calculate the far-field scattering of a single unit-cell at the frequency with minimal reflection, which gives directional scattering. Then we assign different phase patterns and simulate the scattering power pattern of an array of 16 units to mimic the situation shown in Fig.~\ref{fig:array_unit}. The results have a reasonably good agreement with the measured ones in Fig.~\ref{fig:experiment_dyn2} (f), even though in the simulation we assume that the scattering amplitude from each unit is identical and does not change under different modulation phase patterns [see Fig.~\ref{fig:array_unit} (c)].

\section{Experimental measurement}\label{sec:Expmeas}
 
\begin{figure}[t!]
\includegraphics[width=1\columnwidth]{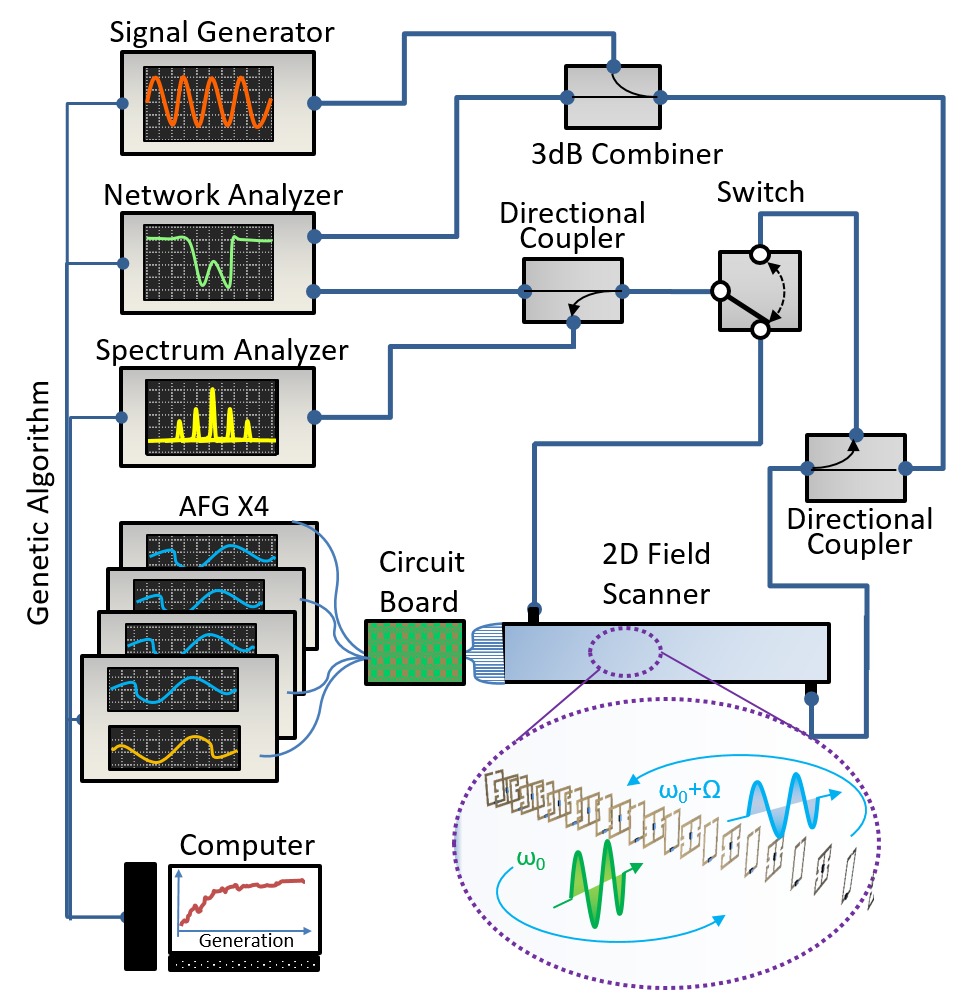}
\caption{Schematic of the  experimental setup for the time-varying Huygens' array. \label{fig:2dwaveguide_setup} }
\end{figure}

\begin{figure}[t!]
\includegraphics[width=0.8\columnwidth]{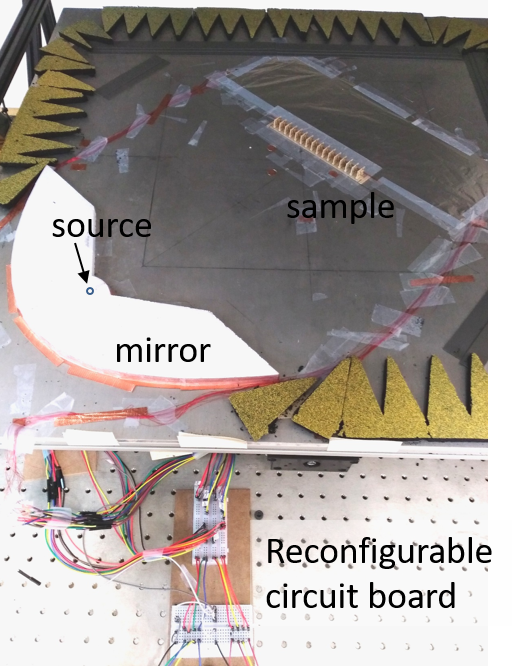}
\caption{Photograph of the experimental platform of the time-varying Huygens' array.  \label{fig:2dwaveguide} }
\end{figure}

To confirm the phase control of sidebands can also be employed in dynamic beam steering, we fabricated an array composed of 16 Huygens' units and positioned the array inside a two dimensional field scanner based on parallel plate waveguide. Each unit has 2 ports (4 patches) connected to the external modulation, as can be seen from Fig.~\ref{fig:array_unit} -- two patches connected to the ground, one patch for the electric meta-atom and one patch for the magnetic meta-atom, and all 64 patches of the array were independently connected with thin enamel wires that connect to the arbitrary function generators. We identified the resonances of the meta-atoms by measuring the scattering in the forward direction with the network analyzer, which shows two dips in the spectrum. We tuned the bias voltages such that the two resonances overlap around 4 GHz when the bias voltages are $U_{E}= U_{M}=-2V$, which were chosen as the starting point of optimization. The array was then excited with a CW signal at 4 GHz ($\lambda\approx 75$mm), which was a collimated beam generated by a parabolic mirror with a monopole source antenna positioned around 315 mm ($\sim 4.2 \lambda$) away from the array center (see Figs.~\ref{fig:2dwaveguide_setup} and \ref{fig:2dwaveguide}). We employed four double-channel arbitrary function generators to provide eight channels of modulation waveforms, four of them were applied on the electric meta-atoms and the other four were used to modulate the magnetic meta-atoms. Due to the limited number of modulation channels available, we only generated modulation phase patterns with 4 discrete phase levels. The patterns were reconfigured by rewiring the connections on a breadboard circuit.

The optimization was performed using the scattered fields measured in the forward ($\theta=0^{\circ}$) and backward ($\theta=180^{\circ}$) directions, under the uniform modulation phase pattern of G1. After optimization, a 2D line scan over a circle centered around the Huygens array was performed to get the scattering pattern. Due to the limited scanning range, the radius of the circle is limited to $2\lambda$. Since the sidebands scattered away from the metasurface decay with distance, the effect of reflection from the parabolic mirror and the non-perfect absorbers at the boundaries is relatively small. The measured level of sidelobes for $\omega_{\pm 1}$ under the phase pattern G1 is around -10 dB. However, for the carrier wave, since the incident beam width is slightly larger than the sample size, the scattered field from the metasurface, the reflections from the parabolic mirror and the non-perfect absorbers at the boundaries interfere with the incident field, which makes an accurate estimation of the spurious scattering quite complicated. Therefore, in the optimization of the array, we only aimed at achieving directive scattering for sidebands $\omega_{\pm 1}$ and maximizing their power ratio over all other sidebands, except for the carrier wave scattered in the backward direction. From the two-dimensional distribution of the measured power under the modulation phase pattern G1, we estimate the levels of the scattered power over the incident one to be around 24\% for sidebands $\omega_{\pm 1}$ and around 3\% for all other higher order sidebands; the rest are the spurious scattering of the carrier wave and the loss due to absorption. 

\section{Genetic algorithm}\label{sec:GA}

\begin{figure}[t!]
\includegraphics[width=1\columnwidth]{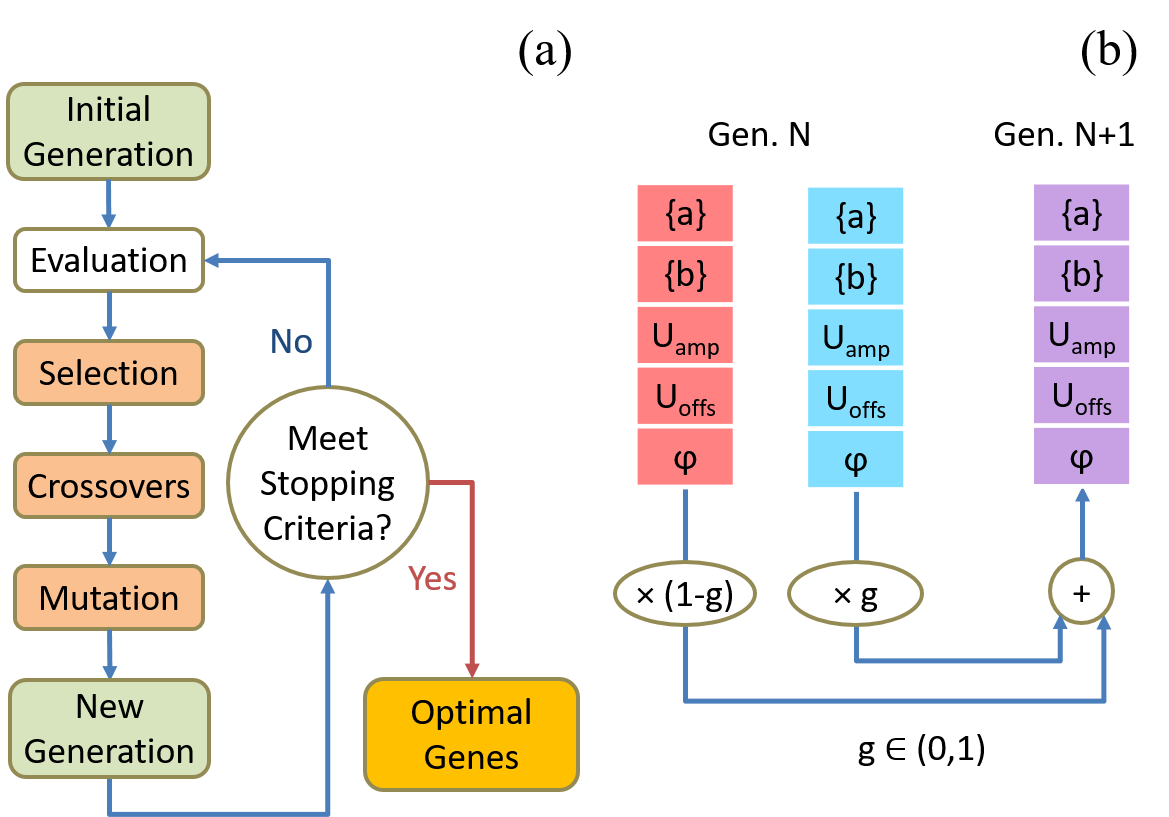}
\caption{ (a) Flow chart of the genetic algorithm. (b) Schematic of the arithmetic-crossover operation.  Two selected parental data sets (chromosomes) from generation N are hybridized to generate a new data set for generation N+1 via convex combination. $g\in (0,1)$ is a random number. \label{fig:genetic} }
\end{figure}

The optimization of the modulation waveforms in this paper was performed via a Genetic Algorithm (GA) \cite{whitley1994genetic,pendharkar2004empirical,haupt2007genetic,mirzaei2014superscattering}. GA is a bio-inspired method that has been widely applied in optimization and search problems. The general procedure of GAs includes several operations that are inspired by evolution and natural selection, including mutation, crossover and selection. Below we will explain the operations in the chart flow of Fig.~\ref{fig:genetic} (a) in the context of waveform optimization used in this paper.

As has been introduced in Sec.~\ref{sec:designHU}, the modulation voltage signal is defined as $\tilde{U}_{\rm E (M)}(t)=U_{\rm E (M), amp}f(t)+U_{\rm E (M), offs}$, with $U_{\rm E (M), amp}$, and $U_{\rm E (M), offs}$ being the modulation amplitude and the DC offset; $f(t)=\sum_{l=1}^{N}a_{\rm E (M)}^{(l)}\cos[l(\Omega t-\varphi_{\rm E (M)})]+b_{\rm E (M)}^{(l)}\sin[l(\Omega t-\varphi_{\rm E (M)})]$ is the normalized waveform constructed using a Fourier series, where we take N=8 as the highest order term. For brevity, we will neglect the subscript of ``E(M)'' below and write the sets of coefficients $a^{(l)}$ and $b^{(l)}$ as $\{a\}$ and $\{b\}$. The problem we need to solve is to find the optimal set of coefficients, i.e. $(\{a\},\{b\},U_{\rm amp}, U_{\rm offs}, \varphi)$, in order to achieve the desired types of sideband scattering.

The flow of optimization, as depicted in Fig.~\ref{fig:genetic} (a), includes many rounds (generations) of operations. In each round, a total number of $N$ sets of coefficients (genes) are generated, and they are used to construct $N$ different modulation signals, which gives $N$ different values of the objective function ($O.F.$). Here we define the objective function as $O.F.=P_{\rm T}\times B_{\rm P}$; $P_{\rm T}$ is the total scattered power of the desired sidebands, and $B_{\rm P}=\mathrm{min}(\zeta_n/\zeta_m,\zeta_m/\zeta_n)$ is a function that gives maximum value when the power of the two desired sidebands are equal. This objective function is employed to maximize the conversion efficiency while maintaining a balanced power when two sidebands are involved.

The best $M$ genes that give the highest values of $O.F.$ are selected as parents to produce $C_M^2=M(M-1)/2$ new genes (offspring) for the next generation via crossovers and mutation. There are many different approaches for crossover operations \cite{pendharkar2004empirical}, such as one-point-crossovers, multi-point-crossovers, uniform-crossovers. Here we employ arithmetic-crossovers that produce offspring via a convex combination of the genes from two parents, as depicted in Fig.~\ref{fig:genetic} (b). After the crossover, mutation is performed by introducing small random values into both the new genes and the old genes to ensure the diversity; then, genes with the lowest values of $O.F.$ are replaced by the best parents (without mutation) and the best ``ancestors'' from previous generations in the new generation. This iterative process continues until the stopping criteria (e.g., the desired value of $O.F.$) are met. In both the simulation and the experiment, we employed 40 genes in each generation, and the results converge well after around 30 to 40 generations.  

The optimized coefficients for the three types of sideband scattering shown in Fig.~\ref{fig:numerical} (d) to (f) are summarized in Table~\ref{table:theory}, where we used the same normalized waveform for electric and magnetic meta-atoms. For double-sideband scattering, we employed symmetric waveforms such that $b^{(l)}=0$.

In the experiment, we set all coefficients to free parameters, and they are summarized in Table~\ref{table:SSF}, ~\ref{table:DSF} ,~\ref{table:DSB} and ~\ref{table:DSB_array}, respectively. We note that $\varphi$ is not an independent parameter since the effect of time-delay can also be accomplished by changing all $a^{(l)}$ and $b^{(l)}$; however, it is much more convenient to use $\varphi$, in particular when we change the relative time-delay of  the two modulation signals while maintaining the waveforms unchanged, as has been shown in Fig.~\ref{fig:experiment_dyn} and ~\ref{fig:experiment_dyn2}.
\begin{table}[t!]
\centering
\caption{Optimized waveform coefficients for the three types of sideband scattering shown in Fig.~\ref{fig:numerical} (d) to (f)}
\label{table:theory}
\bgroup
\def\arraystretch{1.5}
\begin{tabular}{|>{\centering}m{2cm}|>{\centering}m{2cm}|>{\centering}m{2cm}|>{\centering}m{2cm}|}
\hline 
 & Single-sideband forward & Double-sideband forward & Double-sideband bidirectional\tabularnewline
\hline 
\hline 
$a_{{\rm E}}^{(1)}=a_{{\rm M}}^{(1)}$ & 0.357 & -1.121 & -1.186\tabularnewline
\hline 
$a_{{\rm E}}^{(2)}=a_{{\rm M}}^{(2)}$ & 0.166 & 0.055 & 0.002\tabularnewline
\hline 
$a_{{\rm E}}^{(3)}=a_{{\rm M}}^{(3)}$ & -0.124 & 0.215 & 0.297\tabularnewline
\hline 
$a_{{\rm E}}^{(4)}=a_{{\rm M}}^{(4)}$ & -0.159 & -0.037 & 0.010\tabularnewline
\hline 
$a_{{\rm E}}^{(5)}=a_{{\rm M}}^{(5)}$ & -0.033 & -0.068 & -0.131\tabularnewline
\hline 
$a_{{\rm E}}^{(6)}=a_{{\rm M}}^{(6)}$ & 0.073 & 0.004 & -0.004\tabularnewline
\hline 
$a_{{\rm E}}^{(7)}=a_{{\rm M}}^{(7)}$ & 0.063 & 0.040 & 0.070\tabularnewline
\hline 
$a_{{\rm E}}^{(8)}=a_{{\rm M}}^{(8)}$ & 0.005 & 0.003 & 0.002\tabularnewline
\hline
$b_{{\rm E}}^{(1)}=b_{{\rm M}}^{(1)}$ & -0.119 & 0 & 0\tabularnewline
\hline 
$b_{{\rm E}}^{(2)}=b_{{\rm M}}^{(2)}$ & 0.239 & 0 & 0\tabularnewline
\hline 
$b_{{\rm E}}^{(3)}=b_{{\rm M}}^{(3)}$ & 0.173 & 0 & 0\tabularnewline
\hline 
$b_{{\rm E}}^{(4)}=b_{{\rm M}}^{(4)}$ & -0.037 & 0 & 0\tabularnewline
\hline 
$b_{{\rm E}}^{(5)}=b_{{\rm M}}^{(5)}$ & -0.122 & 0 & 0\tabularnewline
\hline 
$b_{{\rm E}}^{(6)}=b_{{\rm M}}^{(6)}$ & -0.068 & 0 & 0\tabularnewline
\hline 
$b_{{\rm E}}^{(7)}=b_{{\rm M}}^{(7)}$ & 0.021 & 0 & 0\tabularnewline
\hline 
$b_{{\rm E}}^{(8)}=b_{{\rm M}}^{(8)}$ & 0.004 & 0 & 0\tabularnewline
\hline
$U_{{\rm E,amp}}$(V) & -2.994 & -0.739 & -0.774\tabularnewline
\hline 
$U_{{\rm E,offs}}$(V) & -2.028 & -2.164 & -1.597\tabularnewline
\hline 
$U_{{\rm M,amp}}$(V) & -2.268 & -0.623 & -0.687\tabularnewline
\hline 
$U_{{\rm M,offs}}$(V) & -1.113 & -1.197 & -1.518\tabularnewline
\hline 
$\varphi_{{\rm M}}-\varphi_{{\rm E}}$($^{\circ}$) & 0 & 0 & -87.6\tabularnewline
\hline 
\end{tabular}
\egroup
\end{table}

\begin{table}[h!]
\centering
\caption{Optimized waveform coefficients for single-sideband forward scattering  shown in Fig.~\ref{fig:experiment} (d)}
\label{table:SSF}
\bgroup
\def\arraystretch{1.5}
\begin{tabular}{|c|c|c|c|c|}
 \hline 
\hline 
$l$ & $a_{{\rm E}}^{(l)}$ & $b_{{\rm E}}^{(l)}$ & $a_{{\rm M}}^{(l)}$ & $b_{{\rm M}}^{(l)}$\tabularnewline
\hline 
1 & -0.372 & 0.161 & -0.345 & 0.213\tabularnewline
\hline 
2 & -0.258 & -0.170 & -0.296 & -0.089\tabularnewline
\hline 
3 & -0.025 & -0.227 & -0.118 & -0.195\tabularnewline
\hline 
4 & 0.113 & -0.080 & 0.051 & -0.129\tabularnewline
\hline 
5 & 0.097 & 0.019 & 0.085 & -0.050\tabularnewline
\hline 
6 & 0.029 & 0.076 & 0.077 & 0.027\tabularnewline
\hline 
7 & -0.018 & 0.034 & 0.019 & 0.034\tabularnewline
\hline 
8 & -0.029 & 0.003 & -0.009 & 0.028\tabularnewline
\hline 
\hline 
$\varphi_{{\rm M}}-\varphi_{{\rm E}}$($^{\circ}$) & $U_{{\rm E,amp}}$(V) & $U_{{\rm E,offs}}$(V) & $U_{{\rm M,amp}}$(V) & $U_{{\rm M,offs}}$(V)\tabularnewline
\hline 
0 & -3.757 & -2.093 & -4.840 & -2.831\tabularnewline
\hline 
\end{tabular}
\egroup 
\end{table}
\begin{table}[h!]
\centering
\caption{Optimized waveform coefficients for double-sideband forward scattering  shown in Fig.~\ref{fig:experiment} (e)}
\label{table:DSF}
\bgroup
\def\arraystretch{1.5}
\begin{tabular}{|c|c|c|c|c|}
 \hline 
\hline 
$l$ & $a_{{\rm E}}^{(l)}$ & $b_{{\rm E}}^{(l)}$ & $a_{{\rm M}}^{(l)}$ & $b_{{\rm M}}^{(l)}$\tabularnewline
\hline 
1 & -0.745 & 0.521 & -0.772 & 0.481\tabularnewline
\hline 
2 & 0.021 & -0.038 & 0.025 & -0.036\tabularnewline
\hline 
3 & -0.009 & -0.067 & 0.001 & -0.068\tabularnewline
\hline 
4 & -0.021 & -0.008 & -0.019 & -0.012\tabularnewline
\hline 
5 & 0.083 & 0.019 & 0.075 & 0.040\tabularnewline
\hline 
6 & -0.023 & 0.009 & -0.024 & 0.001\tabularnewline
\hline 
7 & -0.051 & 0.022 & -0.055 & 0.002\tabularnewline
\hline 
8 & 0.024 & -0.042 & 0.039 & -0.028\tabularnewline
\hline 
\hline 
$\varphi_{{\rm M}}-\varphi_{{\rm E}}$($^{\circ}$) & $U_{{\rm E,amp}}$(V) & $U_{{\rm E,offs}}$(V) & $U_{{\rm M,amp}}$(V) & $U_{{\rm M,offs}}$(V)\tabularnewline
\hline 
0 & -1.142 & -2.424 & -1.686 & -3.128\tabularnewline
\hline 
\end{tabular}
\egroup
\end{table}
\begin{table}[h!]
\centering
\caption{Optimized waveform coefficients for double-sideband bidirectional scattering shown in Fig.~\ref{fig:experiment} (f)}
\label{table:DSB}
\bgroup
\def\arraystretch{1.5}
\begin{tabular}{|c|c|c|c|c|}
\hline 
\hline 
$l$ & $a_{{\rm E}}^{(l)}$ & $b_{{\rm E}}^{(l)}$ & $a_{{\rm M}}^{(l)}$ & $b_{{\rm M}}^{(l)}$\tabularnewline
\hline 
1 & -0.725 & 0.431 & -0.761 & 0.468\tabularnewline
\hline 
2 & -0.110 & 0.203 & 0.003 & -0.135\tabularnewline
\hline 
3 & -0.009 & -0.045 & -0.064 & -0.046\tabularnewline
\hline 
4 & -0.056 & 0.002 & -0.030 & 0.019\tabularnewline
\hline 
5 & 0.050 & -0.046 & -0.032 & -0.104\tabularnewline
\hline 
6 & -0.026 & -0.063 & 0.032 & -0.035\tabularnewline
\hline 
7 & -0.032 & 0.007 & 0.022 & -0.005\tabularnewline
\hline 
8 & 0.019 & 0.008 & 0.039 & 0.033\tabularnewline
\hline 
\hline 
$\varphi_{{\rm M}}-\varphi_{{\rm E}}$($^{\circ}$) & $U_{{\rm E,amp}}$(V) & $U_{{\rm E,offs}}$(V) & $U_{{\rm M,amp}}$(V) & $U_{{\rm M,offs}}$(V)\tabularnewline
\hline 
-89.3 & -1.636 & -1.605 & -1.701 & -3.375\tabularnewline
\hline 
\end{tabular}
\egroup 
\end{table}
\begin{table}[h!]
\centering
\caption{Optimized waveform coefficients for double-sideband unidirectional forward scattering for the array shown in Fig.~\ref{fig:experiment_dyn2} (f) and (g), for plots $\varphi_{\mathrm E}=\varphi_{\mathrm M}$. }
\label{table:DSB_array}
\bgroup
\def\arraystretch{1.5}
\begin{tabular}{|c|c|c|c|c|}
\hline 
\hline 
$l$ & $a_{{\rm E}}^{(l)}$ & $b_{{\rm E}}^{(l)}$ & $a_{{\rm M}}^{(l)}$ & $b_{{\rm M}}^{(l)}$\tabularnewline
\hline 
1&	0.956&	0.988&	1.058&	0.984\tabularnewline
\hline 
2&	-0.021&	-0.003&	0.140&	0.048\tabularnewline
\hline 
3&	-0.085&	0.106&	-0.126&	0.117\tabularnewline
\hline 
4&	0.049&	-0.011&	-0.065&	0.149\tabularnewline
\hline 
5&	-0.013&	0.003&	-0.026&	-0.052\tabularnewline
\hline 
6&	0.076&	-0.068&	-0.019&	0.023\tabularnewline
\hline 
7&	0.009&	0.025&	-0.010&	0.030\tabularnewline
\hline 
8&	-0.027&	0.038&	-0.025&	-0.063\tabularnewline
\hline 
\hline 
$\varphi_{{\rm M}}-\varphi_{{\rm E}}$($^{\circ}$) & $U_{{\rm E,amp}}$(V) & $U_{{\rm E,offs}}$(V) & $U_{{\rm M,amp}}$(V) & $U_{{\rm M,offs}}$(V)\tabularnewline
\hline 
2.072& -4.004	&-2.741&-12.615&	-1.036\tabularnewline
\hline 
\end{tabular}
\egroup 
\end{table}

\section*{References}


%

\end{document}